\documentclass[aps,prx,amsmath,amssymb,twocolumn,superscriptaddress]{revtex4-2}

\usepackage{graphicx}
\usepackage{bm}
\usepackage[dvipsnames]{xcolor}
\usepackage{cancel}
\usepackage[caption=false,font=normalfont]{subfig}
\usepackage{floatrow}
\usepackage{bbold}
\usepackage{ulem}
\usepackage{title_authors}

\usepackage{tikz}
\usetikzlibrary{positioning, shapes.geometric}
\usepackage{pdfpages}
\makeatletter
\AtBeginDocument{\let\LS@rot\@undefined}
\makeatother

\usepackage{microtype}

\usepackage{glossaries}

\usepackage[colorlinks=true,linkcolor=blue,citecolor=red]{hyperref}
\usepackage[capitalise]{cleveref}

\newcommand\TR{\mathcal{T}}

\renewcommand\vec\boldsymbol

\newcommand\outline[1]{[{\color{blue} #1}]}

\renewcommand\outline[1]{}
\let\oldsubsection\subsection
\renewcommand\subsection[1]{\textit{#1} -- }

\newcommand{\be}[0]{\begin{equation}}
\newcommand{\ee}[0]{\end{equation}}

\def\ba#1\ea{\begin{align}#1\end{align}}

\newcommand{\bmat}[0]{\begin{bmatrix}}
\newcommand{\emat}[0]{\end{bmatrix}}

\def\RR{\mathbf{R}}

\begin{document}

\title{\paperTitle}
\paperAuthors

\let\oldaddcontentsline\addcontentsline

\begin{abstract}
In twisted bilayer graphene, a unified understanding of the mechanisms governing temperature-dependent electronic spectra and thermodynamic properties remains controversial despite extensive theoretical efforts. Here, we present a comprehensive theoretical framework that quantitatively accounts for scanning tunneling spectroscopy, quantum twisting microscopy, and thermodynamic properties of magic angle twisted bilayer graphene. We demonstrate that the observed behavior arises from the interplay between electron correlations and external symmetry-breaking induced by strain and lattice relaxation. These effects act cooperatively to shape the emergent electronic behavior, leaving characteristic signatures  across spectroscopy, compressibility and entropy.
\end{abstract}

\maketitle

\begin{figure*}
    \centering
    \includegraphics[width=\textwidth]{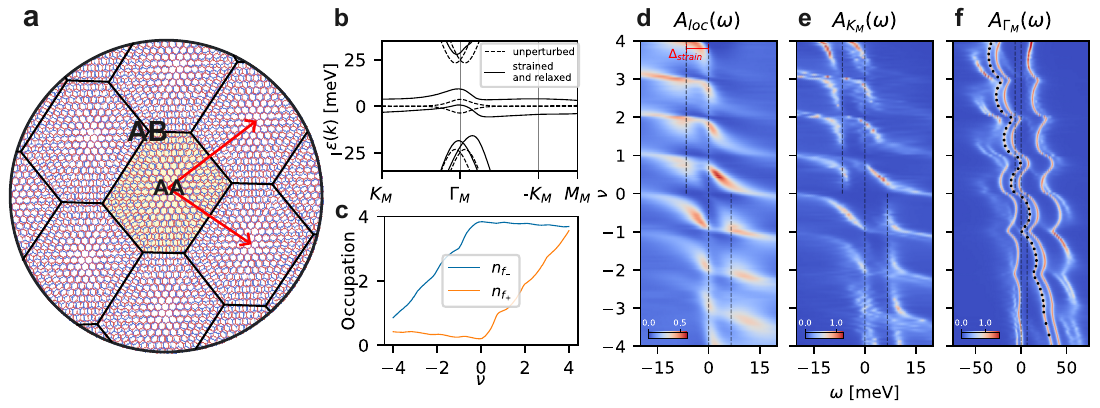}
    \caption{\textbf{a} Sketch of strained twisted bilayer graphene lattice structure. The red arrows represent the moiré lattice vectors, while the shaded area is the strained unit cell. \textbf{b} Non-interacting dispersion of the unperturbed (dashed) and strained and relaxed (solid) THF models for the $K$ valley. \textbf{c} Occupation of the $f_{+}$ and $f_{-}$ orbitals as a function of filling. 
    \textbf{d} Local (mBZ-averaged) spectral function. \textbf{e} Spectral function at the $K_{M}$-point and \textbf{f} at the $\Gamma_{M}$-point for fillings $\nu \in [-4, 4]$ with $0.15\%$ uniaxial heterostrain and lattice relaxation. All data are in the symmetric phase and for $T=11.6$ K. The black dotted line in \textbf{f} is the $\-\mu(\nu)$ curve. The dashed vertical lines are drawn at $\pm M_f \epsilon_-$ which is the magnitude of the strain-induced splitting in the non-interacting model.}
    \label{fig:fig1}
\end{figure*}

 \subsection{Introduction}
Strain and lattice relaxation are ubiquitous effects in graphene moiré systems such as twisted bilayer graphene (TBLG).
Strain asymmetrically stretches the hexagonal moiré unit cell of the system, as sketched in~\cref{fig:fig1}a. TBLG samples typically have substantial heterostrain ($\epsilon = 0.1\%$--$0.4\%$), stemming from several sources during sample preparation and handling, with ultralow-strain samples ($\epsilon<0.1\%$) being rare~\cite{nuckollsQuantumTexturesManybody2023,kerelskyMaximizedElectronInterctions2021,turkelOrderlyDisorder2022,kazmierczakStrainFieldsTwisted2021}. 
Typical and ultralow-strain samples are known to feature different correlated insulating states at low temperatures~\cite{nuckollsQuantumTexturesManybody2023}, highlighting the need for the modeling of electron correlations in the presence of lattice effects~\cite{zhangHeavyFermionsMass2025,caoCorrelatedInsulatorBehaviour2018,nuckollsStronglyCorrelatedChern2021,stepanovCompetingZeroField2021,turkelOrderlyDisorder2022,wongCascadeElectronicTransitions2020,xieSpectroscopicSignaturesManyBody2020,zhangHeavyFermionsMass2025,kazmierczakStrainFieldsTwisted2021}.

Lattice relaxation follows from the energetics of the emergent moir\'{e} pattern, as locally AB (Bernal) regions are energetically favored compared to the locally AA (aligned) regions. 
This results in both in-plane and out-of-plane deformations of the atomistic graphene structure, causing the AA regions to shrink relative to the AB regions and bringing the two layers closer to each other at the AB regions compared to the AA regions~\cite{caoCorrelatedInsulatorBehaviour2018,choiElectronicCorrelationsTwisted2019,haoElectricFieldTunable2024,xieSpectroscopicSignaturesManyBody2020,yuSpinSkyrmionGaps2023}

Both strain and lattice relaxation effects reduce the symmetries of the system. In the absence of perturbations, TBLG is well described by the seminal Bistritzer-MacDonald (BM) continuum model~(\cite{bistritzerMoireBandsTwisted2011} and~\cref{fig:fig1}b) which, in addition to the translational symmetries and $SU(2)$ spin symmetry, possesses $C_{3z}, C_{2x}, C_{2z},\TR$ symmetries and an emergent anti-commuting particle-hole symmetry $P$. 
Heterostrain generically breaks $C_{3z}$ and $C_{2x}$, splitting the flat band manifold, while lattice relaxation breaks $P$~\cite{herzog-arbeitmanHeavyFermionsEfficient2024}, causing the upper flat bands to become more dispersive than the lower flat bands.

In this paper we show that the breaking of some of the lattice and particle-hole symmetries, together with a proper evaluation of electronic correlations, are the key to unlock a unified description of three ubiquitous and hitherto unexplained features observed in TBLG experiments: (i) the emergence of a filling-independent persistent spectral feature at $V_b\sim 10$~meV in the local spectral function~\cite{wongCascadeElectronicTransitions2020, xiaoInteractingEnergyBands2025,zhangHeavyFermionsMass2025}, (ii) a transition from $8$-fold degenerate to $4$-fold degenerate local moments in the flat band manifold when lowering the temperature~\cite{zhangHeavyFermionsMass2025,rozenEntropicEvidencePomeranchuk2021,saitoIsospinPomeranchukEffect2021}, and (iii) the particle-hole asymmetry  of positive vs negative fillings measured from the charge neutrality point (CNP), with  charge compressibility varying more strongly on the electron (e)-doped side~\cite{saitoIsospinPomeranchukEffect2021, rozenEntropicEvidencePomeranchuk2021,dasSymmetryBrokenChern2021,pierceUnconventionalSequenceCorrelated2021,wongCascadeElectronicTransitions2020}, and superconductivity showing higher stability on the hole (h)-doped side~\cite{caoUnconventionalSuperconductivityMagic2018,luSuperconductorsOrbitalMagnets2019,yankowitzTuningSuperconductivityTwisted2019} across many samples.

We find that typical strain values $\sim$$0.1\%$ induce a crystal-field–like splitting of the flat-band manifold of the order of $10$~meV. This directly results in a $\sim$$10$~meV persistent feature in the local density, that matches the filling-independent peak reported in \cite{wongCascadeElectronicTransitions2020, xiaoInteractingEnergyBands2025}. 

Furthermore, our results show that relaxation-induced particle-hole asymmetry in the non-interacting dispersion carries over to the interacting system physics, with stronger variations in the charge compressibility on the electron-doped side compared to the hole-doped side. 

Interestingly, we find that in typically strained samples, when the system is doped away from the charge neutrality point, correlations are largely frozen in the inactive (away from the Fermi level) flat-band sector, and present in the active (crossing the Fermi level) flat-band sector. This has strong implications for the degeneracy of the local moments, as is observed in the entropy as as function of filling, which progressively vanishes at the CNP as the temperature is lowered below the characteristic scale of the strain-splitting, in good agreement with recent experimental data~\cite{zhangHeavyFermionsMass2025}.

\subsection{Model and method} In order to correctly describe the dynamical properties of the (magic angle) MA-TBLG electronic structure, we make use of the Topological Heavy Fermion model (THF)~\cite{songMagicAngleTwistedBilayer2022}, which represents the physics of the flat bands in terms of maximally localized Wannier orbitals accurately mapping the BM continuum model for $1.05^\circ$-twisted bilayer graphene to a generalized periodic Anderson model~\cite{songMagicAngleTwistedBilayer2022}.
Per moir\'{e} unit cell, two localized $f$-orbitals (per spin and valley) represent the maximally localized Wannier functions, which are centered in the AA-stacked regions, transform like $p_x\pm i p_y$ orbitals under the symmetries of TBLG, and capture $\sim96\%$ of the flat band spectral weight. 
Four dispersive $c$-orbitals (per spin and valley) make up most of the two closest remote bands above and below the flat-band manifold and guarantee the correct elementary band representation of the flat bands by virtue of mode hybridization at the $\Gamma_{M}$ point. 
Per spin and valley, the non-interacting Hamiltonian can be written in the form of a $6\times 6$ matrix $H^\eta_{HF}(k)$ acting on the spinor $(f^\eta_\alpha,c^\eta_a)$ with $a=1,2,3,4$ and $\alpha=1,2$ representing the orbitals and $\eta = K, K'$ representing the valleys.

 The THF model can be naturally extended ~\cite{herzog-arbeitmanHeavyFermionsEfficient2024, refsupplement} to describe strain and lattice relaxation effects~\cite{carrRelaxationDomainFormation2018, vafekContinuumEffectiveHamiltonian2023} which are described by local single particle terms. 
Ref. \cite{herzog-arbeitmanHeavyFermionsEfficient2024} shows that the modifications to the band structure due to strain and lattice relaxation can be incorporated---directly from \textit{ab initio} continuum studies \cite{carrRelaxationDomainFormation2018, vafekContinuumEffectiveHamiltonian2023, kangAnalyticalSolutionRelaxed2025}---into the THF model via first-order perturbation theory, resulting in the perturbed Hamiltonian $H^\eta_{HF, \epsilon, \Lambda}$,
\begin{align}
    H^\eta_{THF, \epsilon, \Lambda}(\mathbf{k}) =H^\eta_{THF}(\mathbf{k}) + \delta H^\eta_\epsilon + \delta H^\eta_\Lambda(\mathbf{k}),
\end{align}
where $\delta H^\eta_\epsilon$, $\delta H^\eta_\Lambda(\mathbf{k})$ represent the corrections due to strain and relaxation respectively, $\epsilon$ parametrizes the uniaxial heterostrain, and $\Lambda$ parametrizes the non-local tunneling terms stemming from lattice relaxation. 

The primary effect of $\delta H^\eta_\Lambda$ is the introduction of two orbital-dependent chemical potential terms that shift the two pairs of $c$-electrons relative to each other and to the $f$-electrons, thus breaking $P$-symmetry. The strain term $\delta H_{\epsilon}^\eta$, instead, incites $f$-orbital hybridization causing the flat-band manifold to split. At $0.15\%$ strain, an average value of the typical strain magnitude found in most samples, the split is about $7$~meV. \cref{fig:fig1}(b) shows the typical band structure with strain and lattice relaxation included.
In the following, we will refer to the split narrow bands as the bonding and anti-bonding flat bands, and to the local eigenfunctions of the strained $f$-subspace Hamiltonian (which are eigenstates of $\sigma_y$) as the $f_{-}$ and $f_{+}$ orbitals. 

To model Coulomb interaction in this setup, we assume a double-gate screened interaction with an inter-gate distance $\xi=10$~nm and dielectric constant $\epsilon = 6$. From this, we deduce the value of the many-body interaction coefficients, with the dominant term being the $ff$-interaction $U = 57.95$ meV, and the $cf$ and $cc$ terms circa $15$ meV weaker. Given the width of the narrow bands is of the order of $10$  meV~\cite{songMagicAngleTwistedBilayer2022,herzog-arbeitmanHeavyFermionsEfficient2024,refsupplement}, the interaction-to-bandwidth ratio implies that correlation effects are extremely pronounced. In fact, theoretical simulations show how a mean-field description is not very accurate in capturing spectral and response properties of the system, as well as the temperature scale for phase transitions~\cite{raiDynamicalCorrelationsOrder2024,dattaHeavyQuasiparticlesCascades2023}.
Hence, we employ the charge self-consistent Dynamical Mean-Field Theory (DMFT) framework developed in~\cite{raiDynamicalCorrelationsOrder2024, huSymmetricKondoLattice2023}, making use of the CT-QMC solver \textit{w2dynamics}~\cite{wallerbergerW2dynamicsLocalOne2019} to study the effects of correlations in the presence of strain and lattice relaxation in absence of spontaneous symmetry breaking. We treat the $f$-subspace local interactions at all orders, and all other non-local two-body terms, which fall outside the DMFT approximation, at the Hartree level by coupling to the self-consistently adjusted density matrix. We disregard spontaneous symmetry breaking by setting a temperature of $T \ge 11.6 K$, above the onset of ordering~\cite{raiDynamicalCorrelationsOrder2024}. For more details on the solver and method see~\cite{refsupplement}.

\subsection{Spectral persistent feature and charge sector freezing}

\begin{figure}
    \centering
    \includegraphics[width=\textwidth]{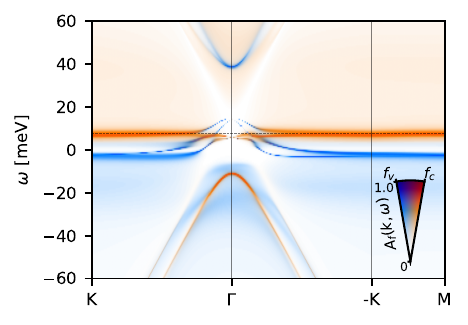}
    \caption{ Momentum-resolved $f$-projected spectral function with lattice relaxation and $0.15\%$ strain at filling $\nu=-0.8$ and $T=11.6 K$, projected onto the strain split $f_{+}$ (orange) and $f_{-}$ (blue) orbitals. On the h-doped side ($\nu<0$), the persistent feature at around $10$~meV is made up of unoccupied $f_{+}$ spectral weight. }
    \label{fig:fig2}
\end{figure}

In Fig.~\ref{fig:fig1}(d-f) we show how the spectral function of TBLG evolves with filling when both strain and lattice relaxation are included. The results presented here refer to a strain level $0.15\%$. The dependence of the spectra on the strain mangitude is discussed in the end matter.
These figures can be compared directly with the local density measured via scanning tunneling microscopy (STM)~\cite{wongCascadeElectronicTransitions2020,kimImagingInterValley2023,nuckollsStronglyCorrelatedChern2021,xieSpectroscopicSignaturesManyBody2020,zhouCoexistenceReconstructedUnreconstructed2023} and the Quantum twisting microscope~\cite{inbarQuantumTwistingMicroscope2023,leeRevealingElectronInteractions2025,xiaoInteractingEnergyBands2025}, as well as the previous DMFT data on unstrained models~\cite{raiDynamicalCorrelationsOrder2024, dattaHeavyQuasiparticlesCascades2023}.
Similarly to the unstrained and unrelaxed system, the spectral function features cascade transitions, with side bands forming patterns that repeat every time the number of electrons per moir\'{e} unit cell changes by an integer, and the $f$- and $c$-electron occupations resetting.
However, the spectra from the perturbed model reproduce one important feature in the experimental data that has thus far eluded explanation.
QTM~\cite{xiaoInteractingEnergyBands2025} and STM~\cite{wongCascadeElectronicTransitions2020} spectra report the presence of a ``persistent" (filling-independent) feature at around $\sim$$10$~meV on the h-doped site and $\sim$$-10$~meV on the e-doped side. 
While this feature was absent in previous unstrained DMFT calculations~\cite{raiDynamicalCorrelationsOrder2024, dattaHeavyQuasiparticlesCascades2023}, our spectral functions of the perturbed model reproduce the persistent feature (cf. the dashed vertical lines in~\cref{fig:fig1}(d-f)) and provide a natural explanation for the corresponding excitations.

\cref{fig:fig1}(d) shows the local spectral function (i.e. averaged over the mini BZ of TBLG), while (e) and (f) are the spectral functions at the $K_{M}$ and $\Gamma_{M}$ high-symmetry points. The theoretical data are in good agreement with the QTM results at the corresponding momenta~\cite{xiaoInteractingEnergyBands2025}. At the $K_{M}$ point, the spectrum presents a well defined gap around the Fermi energy, which is maximum around the CNP. The local spectral function, however, features residual spectral resonance crossing the gap. This is mostly due to the $c$-electron spectral weight at $\Gamma_{M}$, as explained in detail in~\cite{refsupplement}. Since the $c$-electrons are weakly correlated, the associated spectral peaks effectively behave as noninteracting bands, rigidly shifting upon chemical potential variations. This gives rise to the striking correspondence between the central spectral peak and the $-\mu(\nu)$ curve, represented by the black dotted line in ~\cref{fig:fig1}(f), which is also observed experimentally in~\cite{xiaoInteractingEnergyBands2025}.

The breaking of $C_{3z}$ by strain causes the $8$-fold degenerate flat-band manifold in the unperturbed THF to split into two sets of degenerate anti-bonding and bonding $f$-bands.
In~\cref{fig:fig1}(c), we follow the occupation of the $f_{+}$ and $f_{-}$ electrons as a function of total filling $\nu$. We find that on the e-doped side, the $f_{-}$ manifold occupation is almost complete and constant, while the $f_{+}$ electrons are active, and vice versa on the h-doped side.
This suggests that the persistent feature is related to excitations to the inactive sector. Since this switches at the charge neutrality point, it appears at positive or negative bias depending on whether the system is hole- or electron-doped, and the energy at which it appears depends on the magnitude of strain in the sample.

We confirm this hypothesis by looking at the momentum-resolved spectral function in~\cref{fig:fig2}, obtained at $T=11.6K$ and doping $\nu=-0.8$. The hue represents whether the spectral weight comes from the $f_{-}$ (blue) or $f_{+}$ (orange) orbitals.  
Since this is on the hole-doped side of CNP, the occupation of the $f_+$ (orange) electrons is nearly zero. Dynamical correlations are unimportant in this sector and the energy of the corresponding spectral weight is set by the strain splitting. 
This spectral weight is precisely the content of the persistent feature (marked by the dashed horizontal line in ~\cref{fig:fig1}(e,f,g), and at this filling, is composed of $f_{+}$ (orange) electrons. 
The active sector is composed of the $f_-$ (blue) electrons, whose occupation changes as a function of filling. 
This sector exhibits dynamical correlations, developing the Hubbard bands and the zero-bias resonance forming the cascades. At the considered filling, this spectral weight is pinned close to the Fermi level.  

In agreement with the observations in the QTM experiment~\cite{xiaoInteractingEnergyBands2025}, this contribution is absent at the $\Gamma_{M}$ point.
We arrive at the conclusion that strain causes the flat bands to split (already as a single particle effect) into anti-bonding and bonding subsets, only one of which is active depending on whether the system is electron- or hole-doped. The inactive sector causes the filling-independent persistent feature, while the active sector is responsible for the correlations and cascade resets.

\subsection{$P$-symmetry breaking and inverse compressibility}

\begin{figure}
    \centering
    \includegraphics[width=\linewidth]{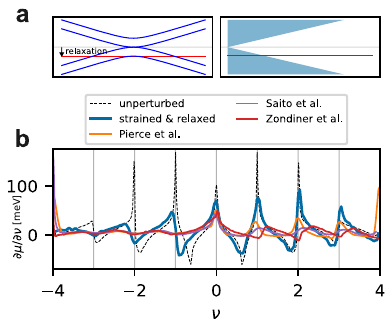}
    \caption{\textbf{a} Sketch of the bare (zero-hybridization) dispersion of the $f$- and $c$-electrons in the relaxed THF model (left) and the corresponding density of states (right),  \textbf{b} Inverse charge compressibility $\partial\mu/\partial\nu$ as a function fo filling in TBLG with and without lattice relaxation and $0.15\%$ strain and $T=11.6$ K, compared with experimental data~\cite{pierceUnconventionalSequenceCorrelated2021, zondinerCascadePhaseTransitions2020, saitoIsospinPomeranchukEffect2021}. P-symmetry breaking due to lattice relaxation results in stronger peaks and troughs in the inverse compressibility on the electron-doped side.}

    \label{fig:fig3}
\end{figure}

At very low T, symmetry-broken integer-filling insulators are more stable on the electron side~\cite{stepanovUntyingInsulatingSuperconducting2020,saitoIndependentSuperconductorsCorrelated2020}.
Lattice relaxation effects are responsible for particle-hole symmetry breaking, as Hartree-Fock studies have shown~\cite{kwanKekuleSpiralOrder2021,wagnerGlobalPhaseDiagram2022}. Yet, mean-field solutions are unable to correctly capture the sawtooth behavior of compressibility and the negative compressibility regions~\cite{raiDynamicalCorrelationsOrder2024,saitoIsospinPomeranchukEffect2021,zondinerCascadePhaseTransitions2020,pierceUnconventionalSequenceCorrelated2021}.

To explicitly quantify the impact of lattice relaxation induced $P$-symmetry breaking on theormodynamic observables, we study 
the relation between the chemical potential and the total filling.~\cref{fig:fig2}(b) shows the inverse charge compressibility, $\partial\mu/\partial\nu$, of the strained and relaxed THF model compared to the unperturbed THF model from ~\cite{raiDynamicalCorrelationsOrder2024} and experimental data~\cite{zondinerCascadePhaseTransitions2020,pierceUnconventionalSequenceCorrelated2021,saitoIsospinPomeranchukEffect2021}.

We find that the addition of strain and lattice relaxation terms allows us to capture an important qualitative feature missing in the unperturbed THF model. Multiple experiments~\cite{zondinerCascadePhaseTransitions2020,pierceUnconventionalSequenceCorrelated2021,saitoIsospinPomeranchukEffect2021} across different samples consistently find a strong particle-hole asymmetry: the variations in the charge compressibility are markedly stronger on the e-doped side compared to the h-doped side. The THF model with strain and lattice relaxation reproduces this phenomenon, and it can be understood by considering the primary effect of the lattice relaxation part of the Hamiltonian, which is a relative downward shift of the $f$-electron energies with respect to the charge neutrality point of the $c$ Dirac bands. This is shown schematically in Fig.~\ref{fig:fig3}(a) where the bare $f$ dispersion is represented by the red line, shifted down with respect to the touching point of the $c$ bands. As a consequence, there is a higher $c$ density of states available on the hole-doped side compared to the electron-doped side, and therefore there is stronger hybridization on the hole-doped side compared to the electron-doped side. This suppresses the local moment on the hole-doped side relative to the electron-doped side, which results in the softening of the compressibility variations. This is consistent with Hartree-Fock simulations which find that the ordered insulators on the electron-doped side have a larger gap than the ones on the hole-doped side~\cite{kwanKekuleSpiralOrder2021,wagnerGlobalPhaseDiagram2022}. Note that one might naively expect that since the upper flat band is more dispersive, correlations would be stronger on the hole-doped side. However, the crucial factor affecting the strength of correlations is the hybridization, which in turn correlates positively with the higher $c$ density of states available on the hole-doped side.

Another key improvement with respect to previous simulations regards the position and shape of the inverse compressibility maxima. Though different experimental results show variations in the position and relative size of the peaks for electron doping (positive $\nu$), the behavior for hole doping (negative $\nu$) is remarkably consistent~\cite{saitoIsospinPomeranchukEffect2021,rozenEntropicEvidencePomeranchuk2021,pierceUnconventionalSequenceCorrelated2021,zondinerCascadePhaseTransitions2020}, and shows a marked depinning of the inverse compressibility maxima from integer total occupations.
The inclusion of strain and relaxation terms immediately leads to a better estimation of the position and size of the compressibility maxima on the hole-doping side (blue solid line) with respect to the unstrained model (dashed black line in~\cref{fig:fig3}b).
The increased hybridization on the hole-doped side again plays a crucial role in the depinning. The hybridization has a smoothening effect making the $f$-sector less quantum-dot like, lowering the penalty of non-integer occupation. 
In other words, the system is more metallic on the hole-doped side than on the electron-doped side (see further discussion in the supplemental material~\cite{refsupplement}).

\subsection{Entropy} 
Entropy has been recently used to experimentally assess the presence and degeneracy of local  moments in TBLG ~\cite{zhangHeavyFermionsMass2025}. We calculate the entropy as a function of filling, for different values of temperature, by means of the Maxwell relation
\begin{equation}
    S(\nu,T)=-\int_{4}^{\nu}\bigg(\dfrac{\partial \mu}{\partial T}\bigg)_{\nu'} \mathrm{d}\nu'
\end{equation}
where we assume that the band insulator at full filling ($\nu=4$) has zero entropy. 

\begin{figure}
    \centering
    \includegraphics[width=\linewidth]{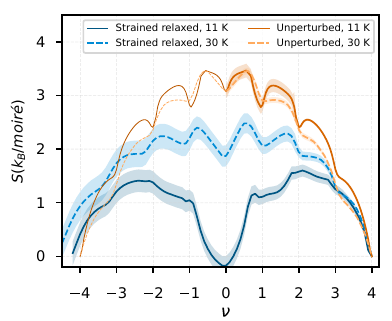}
    \caption{Entropy as a function of filling of unperturbed TBLG (orange lines) compared to that of strained ($0.15\%$) and relaxed TBLG (blue lines). 
    The error bars (shaded area) are obtained by propagating the relative mean square deviation of a sample of DMFT-CTQMC iterations after convergence. In the unstrained case, the entropy on the hole-doping side is not calculated directly, since it is is determined by $P$-symmetry. It is denoted by thin lines without error bars.}
    \label{fig:fig4}
\end{figure}

~\cref{fig:fig4} shows the entropy of TBG with and without strain and relaxation at two temperatures. 
At $11$~K, the entropy curves with and without strain are qualitatively different.
In the unstrained case, the entropy curve oscillates around a rough semi-circular envelope spanning $\nu\in[-4,4]$ with a maximum at the charge neutrality point.
In contrast, in the strained case, the entropy goes to zero at the charge neutrality point, with two semi-circular envelopes spanning $\nu\in[-4,0]$ and $\nu\in[0,4]$.
This can be explained by the splitting of the $8$-fold degenerate flat-band manifold into two $4$-fold degenerate anti-bonding and bonding bands by strain. 
The vanishing entropy at the charge neutrality point is a consequence of the gap between the anti-bonding and bonding flat bands. 
At higher temperatures ($30$~K), thermal excitation can overcome the strain gap restoring the $8$-fold degeneracy of the local moments. 
The degeneracy is reflected in the fine features of the entropy curve. There are local maxima between neighboring integer fillings giving rise to an overall $8$-lobed structure.
Consequently, at high temperatures, both unstrained and strained data have the same qualitative features: $8$ lobes and finite entropy at CNP. The shape of the entropy curves match favorably with experimental measurements~\cite{zhangHeavyFermionsMass2025,rozenEntropicEvidencePomeranchuk2021}.

\subsection{Discussion}
The spectrosopic $\pm 10$meV feature, the evolution of entropy with filling and temperature and the pronounced particle–hole asymmetry observed, for example, in charge compressibility measurements constitute key experimental constraints for a proper theoretical description of twisted bilayer graphene. 
Existing theoretical efforts have partially addressed this set of phenomena using a broad range of frameworks, including low-energy Dirac semimetals~\cite{pierceUnconventionalSequenceCorrelated2021,rozenEntropicEvidencePomeranchuk2021},  interaction-induced band renormalizations~\cite{rademakerChargeSmootheningBand2019, zhuWeakCouplingTheory2024}, lifting of flavor degeneracies via spontaneous symmetry breaking or external perturbations~\cite{xieWeakFieldHall2021,bultinckGroundStateHidden2020}, heavy-fermion and  mixed-valence scenarios~\cite{zhangHeavyFermionsMass2025,lauTopologicalMixedValence2025}, incipient subband formation in paradigmatic Hubbard models~\cite{wongCascadeElectronicTransitions2020,ghoshThermopowerProbesEmergent2025}, and exotic trion excitations~\cite{zhaoTopologicalMottLocalization2025,ledwith2025nonlocalmomentschernbands}. However, a single microscopic framework capable of consistently accounting for all these effects has remained elusive.

Here, we showed that these seemingly distinct experimental signatures can only be understood by simultaneously accounting for correlations, strain and lattice-relaxation effects inherent to realistic twisted bilayer graphene. We identified strain-induced splitting of the flat-band manifold as the central mechanism underlying both the emergence of a persistent feature in the local spectral density and the crossover from an eightfold-degenerate to a fourfold-degenerate local moment upon lowering the temperature~\cite{zhangHeavyFermionsMass2025}. At low temperatures, one of the two strain-split sectors becomes effectively inactive and is energetically positioned so as to account quantitatively for the persistent ±10 meV feature observed in QTM and STM experiments. The freezing of this inactive sector naturally explains the reduction of the local-moment degeneracy, and our numerical estimates of the resulting entropy are in excellent agreement with recent experimental measurements.

The identification of an inactive charge sector also has important implications for future theoretical modeling of correlated states in twisted bilayer graphene~\cite{songKondoPhaseTwisted2024}. Any faithful low-energy theory must incorporate the effective freezing of one sector upon lowering the temperature and collective instabilities—most prominently superconductivity—must be understood as arising within this constrained low-energy landscape. Finally, lattice relaxation and the associated breaking of particle–hole symmetry render the bonding flat bands more dispersive than the antibonding flat bands; this asymmetry is inherited by the interacting system and naturally leads to stronger correlation effects on the electron-doped side compared to the hole-doped side. More broadly, this work underscores the exceptional sensitivity of correlated quantum materials to externally imposed symmetry breaking, revealing a general route toward controlling collective electronic behavior.

\subsection{Data availability}
The data that support the findings of this work will be deposited in Zenodo and made publicly available upon publication.

\subsection{Acknowledgements} We thank Oskar Vafek, Erez Berg and Shahal Ilani for useful discussions. L.C., G.R.,  and T.W. acknowledge support from the Cluster of Excellence ‘CUI: Advanced Imaging of Matter' – EXC 2056 (Project No. 390715994), and SPP 2244 (WE 5342/5-1 project No. 422707584). L.C. gratefully acknowledges the scientific support and HPC resources provided by the Erlangen National High Performance Computing Center (NHR@FAU) of the Friedrich-Alexander-Universität Erlangen-Nürnberg (FAU) under the NHR project b158cb. G.R., L.C., R.V., G.S., and T.W. acknowledge support from the Deutsche Forschungsgemeinschaft (DFG, German Research Foundation) through QUAST FOR 5249 (Project No. 449872909, projects P4 and P5). G.S. acknowledges financial support by the Deutsche Forschungsgemeinschaft (DFG, German Research Foundation) under Germany's Excellence Strategy–EXC2147 ``ct.qmat'' (project‐id 390858490). R.V. thanks the Deutsche Forschungsgemeinschaft (DFG, German Research Foundation) through the TRR 288 - 422213477 and project Nr. VA 117/23-1 — 509751747. B.A.B. and H. H. were supported by the Gordon and Betty Moore Foundation through Grant No. GBMF8685 towards the Princeton theory program, the Gordon and Betty Moore Foundation’s EPiQS Initiative (Grant No. GBMF11070), the Office of Naval Research (ONR Grant No. N00014-20-1-2303), the Global Collaborative Network Grant at Princeton University, the Simons Investigator Grant No. 404513, the NSF-MERSEC (Grant No. MERSEC DMR 2011750), Simons Collaboration on New Frontiers in Superconductivity (SFI-MPS- NFS-00006741-01), the Schmidt Foundation at Princeton University and  the Princeton Catalyst Initiative. H. H. and D. C. also acknowledge support from the European Research Council (ERC) under the European Union’s Horizon 2020 research and innovation program (Grant Agreement No. 101020833).

\renewcommand{\addcontentsline}[3]{}
\bibliography{bibliography_sundry_byhand,bibliography_strainrelaxation}

\onecolumngrid
\begin{center}
	\textbf{\large End Matter}\\[.2cm]
\end{center}

\begin{figure}[ht]
    \centering
    \includegraphics[width=\linewidth]{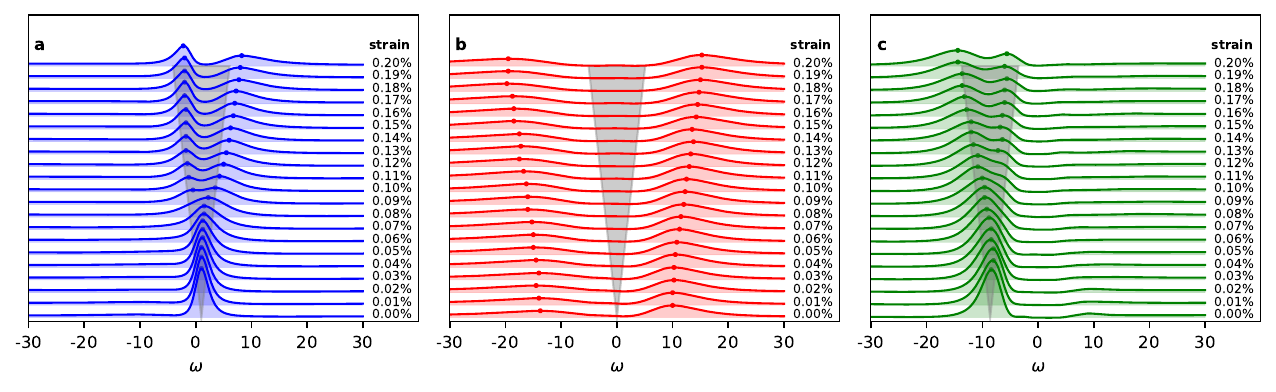}
    \caption{Development of the strain-induced resonance splitting in the local spectral function for three values of doping: \textbf{(a)} $\nu = -0.8$, fractional filling, where in the unstrained case a pinned resonance at the Fermi level is observed,  \textbf{(b)} $\nu = -0.0$, the charge neutrality point, where two Hubbard peaks are observed, and \textbf{(c)} $\nu = 2.0$, where in the unstrained case a coherent spectral peak is present at $\omega\approx -10 meV$. The grey shaded area represents the energy splitting of the $f$-orbitals in the noninteracting model. These have been shifted to be centered around the frequency of the noniteracting spectral maximum in panels \textbf{(a,c)} so as to offer a comparison between the interacting and noninteracting splitting magnitude.}
    \label{fig:endmatter}
\end{figure}

As discussed in the main text, the splitting of the $f$-band manifold is not consequent from spontaneous symmetry breaking, i.e. local isospin moments ordering.
It is instead due to an ``external" term in the hamiltonian, the effect of which is akin to the crystal field splitting commonly found e.g. in transition metal oxides~\cite{betheTermaufspaltungInKristallen1929,castellaniMagneticStructureV2031978,poteryaevEnhancedCrystalField2007}. Essentially, due to crystal deformations which lower the symmetry of the system, already at the noninteracting level the barycenter of different flat bands is shifted to positive/negative energies respectively. We give a representation of this effect and its consequences in Fig.~\ref{fig:endmatter}, where the local interacting spectral function of the system is plotted, for increasing values of strain, at three different dopings. The considered strain percentage increases from $0$ (unstrained case) to $0.2\%$, a range in which most of the experimental realizations fall~\cite{nuckollsQuantumTexturesManybody2023,kerelskyMaximizedElectronInterctions2021,turkelOrderlyDisorder2022,kazmierczakStrainFieldsTwisted2021,zhangHeavyFermionsMass2025,xiaoInteractingEnergyBands2025}. 
The non-interacting crystal field splitting, symmetric with respect to the Fermi level and linear in strain, is represented  by the grey shaded area. Once electron interactions are included in the pictures, the splitting is renormalized.
By comparing the position of the spectral maxima in Fig.~\ref{fig:endmatter} to the noninteracting one notices that the effect of electron-electron interactions is twofold: first, the overall center of the low-energy manifold may be shifted from zero frequency. This is purely a consequence of interaction, and can be observed already in the unstrained case (see also~\cite{raiDynamicalCorrelationsOrder2024}).
On top of this, a splitting proportional to the noninteracting crystal-field splitting develops. The two peaks shift asymmetrically with respect to the spectral maximum in the unstrained case. In particular, the peak with lower-in-modulus energy tends to shift less away from zero bias than the other. This is reminiscent of the resonance pinning observed in twisted bi- and trilayer graphene, especially at fractional fillings~\cite{kimResolvingIntervalleyGaps2025}.


\let\addcontentsline\oldaddcontentsline
\let\subsection\oldsubsection

\renewcommand{\thesection}{S\arabic{section}}
\renewcommand{\thetable}{S\arabic{table}}
\renewcommand{\thefigure}{S\arabic{figure}}
\renewcommand{\theequation}{S\arabic{section}.\arabic{equation}}

\setcounter{section}{0}                      
\setcounter{table}{0}
\setcounter{figure}{0}
\setcounter{equation}{0}

\onecolumngrid
\pagebreak
\thispagestyle{empty}
\newpage

\begin{center}
	\textbf{\large Supplemental material for \paperTitle}\\[.2cm]
\end{center}

\tableofcontents
\let\oldaddcontentsline\addcontentsline

\section{Strained and relaxed THF Model for TBLG}

\subsection{Noninteracting Hamiltonian}

The noninteracting Hamiltonian of the THF model, as originally derived in~\cite{songMagicAngleTwistedBilayer2022}, has the form
\begin{equation}
    H_{THF}=H^{ff}+H^{cc}+H^{fc}+h.c.
\end{equation}
 The first term is a $8\times 8$ matrix in combined orbital, valley and spin space. The extremely weak inter-$f$ orbital coupling, of the order of $0.1meV$, render this term effectively negligible at the magic angle.

The $fc$ and $cc$ terms can be cumulatively written as

\begin{equation}
    \overbrace{\sum_{|\mathbf{k}|<\Lambda_c}\sum_{aa'\eta\sigma}H_{aa'}^{(c,\eta)}(\mathbf{k})c^\dagger_{\mathbf{k}a\eta\sigma}c_{\mathbf{k}a'\eta\sigma}}^{\hat{H}_{cc}}\nonumber+\underbrace{\frac{1}{\sqrt{N_M}}\sum_{|\mathbf{k}|<\Lambda_c, \mathbf{R}}\sum_{a\alpha\eta\sigma}\left[e^{i\mathbf{k}\cdot\mathbf{R} - \frac{|\mathbf{k}|^2\lambda^2}{2}}H_{a\alpha}^{(fc,\eta)}(\mathbf{k})f^\dagger_ {\mathbf{R}\alpha\eta\sigma}c_{\mathbf{k}a\eta\sigma} + h.c.\right]}_{\hat{H}_{fc}}.
    \label{app:eqn:sp_ham_thf}
\end{equation}
where
\begin{align}
    H^{(c,\eta)}(\mathbf{k}) = \begin{pmatrix}
        0_{2\times 2} & v(\eta k_x \sigma_0 + i k_y \sigma_z)\\
        v(\eta k_x \sigma_0 - i k_y \sigma_z) & M\sigma_x
    \end{pmatrix}, \label{app:eqn:H_c}
\end{align}

and 

 \begin{align}
H^{(fc,\eta)}(\mathbf{k}) = \begin{pmatrix}
    \gamma\sigma_0 + v'(\eta k_x\sigma_x + k_y\sigma_y) & 0_{2\times 2}
\end{pmatrix}, \label{app:eqn:H_fc}
\end{align}
where $\eta=\pm1$ is the valley index, and the Pauli matrices are in orbital space. The Hamiltonian is identical per spin. The noninteracting parameters of the model are the following, for a twist angle of $\theta=1.05^{\circ}$: 

\begin{equation*}
    \gamma = -24.8 meV,\quad M = 3.7 meV, \quad v = -4.3 eV\cdot\AA, \quad v'=1.6 eV\cdot \AA
\end{equation*}

To this Hamiltonian we add the following terms, a subset of those derived in~\cite{herzog-arbeitmanHeavyFermionsEfficient2024}, which account for uniaxial heterostrain 

\begin{equation}
H^{\eta}_\epsilon\!=\!\!\!\left( \!\!
\begin{array}{c|cc} 
    M_f  \eta\epsilon_{-}\sigma_{2}& -i\eta\gamma'\epsilon_{+}\sigma_{3} & ic''\eta\epsilon_{-}\sigma_{3}\\ 
    \hline
    h.c.  & c\eta\epsilon_{-}\sigma_{2} & -c'\eta\epsilon_{-}\sigma_{2} \\
    h.c. & h.c. & M'\eta\epsilon_{+}\sigma_2
\end{array}
\! \! \right)
    \label{app:eqn:H_strain}  
\end{equation}

for $\eta=\pm$ valley index and with noninteracting parameters

\begin{equation*}
c = -8750 meV,\quad
c' = 2050 meV,\quad
c'' = -3362 meV,\quad
M_{f} = 4380 meV,\quad
\gamma' = -3352 meV,\quad
M' = -4580 meV
\end{equation*}
and lattice relaxation 

\begin{equation}
H^{\eta}_\Lambda\!=\!\!\!\left( \!\!
\begin{array}{c|cc} 
    \mu_{f}\sigma_{0}& 0 & v_{2}\mathbf{k}\cdot(\sigma_{0},-i\sigma_{3})\\ 
    \hline
    h.c.  & \mu_{1}\sigma_{0} & v_{1}\mathbf{k}\cdot\mathbf{\sigma}^{\star} \\
    h.c. & h.c. & \mu_{2}\sigma_{0}
\end{array}
\! \! \right)
    \label{app:eqn:H_relaxation}  
\end{equation}

with noninteracting parameters

\begin{equation*}
\mu_{f}=0.0 meV,\quad
\mu_{1}= 14.4 meV,\quad
\mu_{2} = 4.5 meV,\quad
v_{1} = 0.2 eV\cdot \AA,\quad
v_{2} = -0.4 eV \cdot \AA.
\end{equation*}

The coefficients $\epsilon_{\pm}$ represent the isotropic and anisotropic heterostrain, which transform real-space coordinates as $\mathbf{r}\rightarrow(1+\varepsilon)\mathbf{r}$ where

\begin{equation}
    \varepsilon=\begin{pmatrix}
        \epsilon_{+}+\epsilon_{-} & 0 \\
        0 & \epsilon_{+}-\epsilon_{-}  \\
    \end{pmatrix}
\end{equation}

We define $\epsilon_{\pm}$ as $\pm(v_{G\mp 1})\epsilon/2$, where $v_{G}=0.16$ is the Poisson ratio for graphene, determining the elongation along one direction when compression is applied along the other, and $\epsilon=0.0015=0.15\%$ is the strain percentage.

\subsection{BZ sampling}

\begin{figure}
    \includegraphics[width=0.5\linewidth]{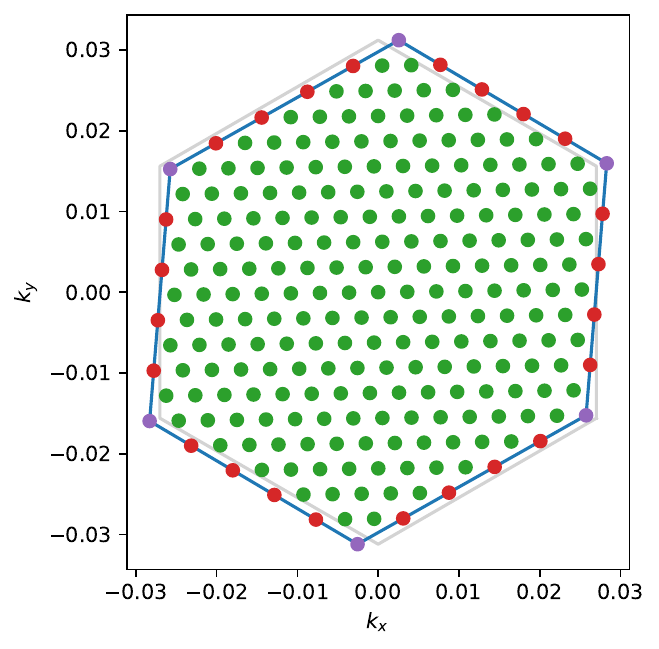}
    \caption{Regular centered sampling of the first moir\'e Brillouin zone for the strained model. The grey contour represents the unstraind mBZ. Edges and corners are weighted $1/2$ and $1/3$ respectively, and represented here by differently colored dots.}
    \label{fig:BZ_sampling}
\end{figure}

The THF Hamiltonian involves a sum over moir\'e lattice sites and and momenta alike. We consider a single lattice site labeled by $\mathbf{R}=(0,0)$, and restrict our $k$-summation on the first mBZ, sampled with a regular centered grid as in~\cite{raiDynamicalCorrelationsOrder2024}. The $k$-points on the edges and corners of the hexagonal BZ are degenerate, and are sampled with a relative weight of $1/2$ and $1/3$ respectively. We then strain the obtained $k$-point mesh with the transformation
\begin{equation}
    \mathbf{k}\rightarrow \mathbf{k} - \sum_{i}(\mathbf{k}\cdot\mathbf{a}_{i})\begin{pmatrix}
        -\epsilon & 0\\
        0 & v_{G}\epsilon\\
    \end{pmatrix}(\mathbf{K}_{3}-\mathbf{K}_{i}).
\end{equation}
Here, $\mathbf{a}_{i}$ are the moiré Bravais lattice vectors and and $\mathbf{K}_{i}$ are obtained by applying the rotation $C_{3}$ $i-1$ times to the monolayer graphene $\mathbf{K}$ point.
See Fig.~\ref{fig:BZ_sampling} for an example grid.  

\subsection{Interacting Hamiltonian}\label{sec:InteractingH}

The interaction Hamiltonian is obtained from the Coulomb integrals
\begin{align}
    \hat{H}_I =\frac{1}{2}\int d^2 \mathbf{r}_1 d^2 \mathbf{r}_2 V(\mathbf{r}_1 - \mathbf{r}_2) : \hat{\rho}(\mathbf{r}_1)::\hat{\rho}(\mathbf{r}_2):,
\end{align}
\noindent where V is the double-gate screened Coulomb potential, for a setup with inter-gate distance $\approx 10$~nm and dielectric constant $\epsilon\approx 6$. 
The $:\rho:$ notation denotes the density evaluated with respect to that at the charge nautrality point, and implies $\mu_{CNP}=0$ for the symmetric model.
Among the two-body operators compatible with the symmetries of the THF model, we consider the following subset to make up the interaction Hamiltonian:

\begin{align}
    \hat{H}_I \approx \hat{H}_U + \hat{H}_W + \hat{H}_V.
\end{align}

These are density-density interactions acting within the $f$ subspace, between $f$ and $c$ and within $c$ respectively. The remaining terms, such as the $f-c$ exchange interaction, have been neglected since they do not qualitatively alter the resulting picture in the symmetric phase~\cite{raiDynamicalCorrelationsOrder2024}.
We define the density operators

\begin{align}
    :f^\dagger_{\mathbf{R}\alpha_1\eta_1\sigma_1} f_{\mathbf{R}\alpha_2\eta_2\sigma_2}: &= f^\dagger_{\mathbf{R}\alpha_1\eta_1\sigma_1} f_{\mathbf{R}\alpha_2\eta_2\sigma_2} - \frac{1}{2}\delta_{\alpha_1\eta_1\sigma_1;\alpha_2\eta_2\sigma_2},\\
    :c^\dagger_{\mathbf{k_1}a_1\eta_1\sigma_1} c_{\mathbf{k_2}a_2\eta_2\sigma_2}: &= c^\dagger_{\mathbf{k_1}a_1\eta_1\sigma_1} c_{\mathbf{k_2}a_2\eta_2\sigma_2} - \frac{1}{2}\delta_{\mathbf{k_1}\alpha_1\eta_1\sigma_1;\mathbf{k_2}\alpha_2\eta_2\sigma_2}.
\end{align}
The (partial) traces over orbital, valley and spin indices of the above operators give the $f$ and $c$ occupations with respect to CNP (dopings) $\nu_{f,(\alpha\nu\sigma)}$ and $\nu_{c,(\alpha\nu\sigma)}$.

In the charge self-consistent DMFT approach, different interaction terms are treated with different degrees of approximation. The on-site Hubbard term 

\begin{align}
    \hat{H}_U = \frac{U}{2}\sum_{\mathbf{R}}\sum_{(\alpha\eta\sigma)\neq (\alpha'\eta' \sigma')}f^\dagger_{\mathbf{R}\alpha\eta\sigma}f_{\mathbf{R}\alpha\eta\sigma}f^\dagger_{\mathbf{R}\alpha'\eta'\sigma'}f_{\mathbf{R}\alpha'\eta'\sigma'}
    - 3.5 U \sum_{\mathbf{R}}\sum_{\alpha\eta\sigma} f^\dagger_{\mathbf{R}\alpha\eta\sigma}f_{\mathbf{R}\alpha\eta\sigma} + \mathcal{O}(1),
\end{align}

 with $U=57.95 meV$, is treated at all-order within DMFT. By constrast, $\hat{H}_{W}$ and $\hat{H}_{V}$ are decoupled at the Hartree level by approximating them as

\begin{align}
    \hat{H}_V^{MF} = V \nu_c \sum_{|\mathbf{k}|<\Lambda,a\eta\sigma} c^\dagger_{\mathbf{k}a\eta\sigma}c_{\mathbf{k}a\eta\sigma} + \mathcal{O}(1),
\end{align}

and

\begin{align}
    \sum_{|\mathbf{k}|<\Lambda_c}\sum_{a\eta\sigma}W_a\nu_f:c^\dagger_{\mathbf{k}a\eta\sigma}c_{\mathbf{k}a\eta\sigma}: +     
\sum_{\mathbf{R}}\sum_{a\alpha\eta\sigma}W_a\nu_{c,a}:f^\dagger_{\mathbf{R}\alpha\eta\sigma}f_{\mathbf{R}\alpha\eta\sigma}: + \mathcal{O}(1).
\end{align}

We assume for simplicity the same interaction coefficient for the $\Gamma_{3}$ and $\Gamma_{1}\oplus\Gamma_{2}$ dispersive orbitals, so that $W_{a}=W=47.12 meV$, and $V=48.33 meV$. These assumptions are consistent with the study of the unstrained and unrelaxed model in the symmetric phase in~\cite{raiDynamicalCorrelationsOrder2024, huSymmetricKondoLattice2023}, and do not alter the qualitative picture.

\section{$f-c$ hybridization function in presence of strain and relaxation}\label{sec:HybFunc}
In the following, we discuss how the strain and relaxation terms modify the hybridization function for the $f$-electrons of the THF model for TBG. 
As a quick reminder of the notation, we start from the 6$\times$6 Hamiltonian for the first valley written in terms of the 2$\times$2 $H^{ff}(\boldsymbol{k})$-block, the 4$\times$4 $H^{cc}(\boldsymbol{k})$-block and the corresponding rectangles connecting  the two subspaces: 
\begin{equation} \label{eq:Hfull}
\begin{pmatrix} 
H^{ff}(\boldsymbol{k}) & H^{fc}(\boldsymbol{k}) \\
\\
H^{cf}(\boldsymbol{k}) & H^{cc}(\boldsymbol{k})
\end{pmatrix}.
\end{equation}
Without strain and relaxation, all four $ff$-elements vanish identically, while in the presence of strain and relaxation they can be nonzero but are still momentum-independent. Therefore, in the following, we keep $H^{ff}$ as a 2$\times$2-matrix explicitly in the calculations but omit its momentum index $\boldsymbol{k}$.
The other blocks are obtained from Eq.~\ref{app:eqn:H_c} and ~\ref{app:eqn:H_fc} for $\eta=1$. 

The hybridization function is a 2$\times$2-matrix (for each valley)
which requires the extraction of local quantities (i.e. summations over all momenta $\boldsymbol{k}$), as well as the projection onto the $ff$-subspace, operations indicated in the following with the subscript ``loc'' and with $\left.  \right|_{ff}$ , respectively.
In terms of the fermionic Matsubara frequency $i\omega$, the hybridization function can be written as
\begin{equation} \label{eq:hyb_fun}
\Delta^{ff}(i\omega) = i\omega \mathbb{I}_{2\times2} - H^{ff}_\text{loc} -  \Big[ \left. G^0_\text{loc}(i\omega) \right|_{ff} \Big]^{-1},
\end{equation}
where $H^{ff}_\text{loc}$ in our case is simply $H^{ff}$ because the latter is independent of $\boldsymbol{k}$ and the sum is normalized to 1.
To see the effect of the $ff$-projection on the non-interacting Green's function, we first express 
$G^0_\text{loc}(i\omega)$ in the full $c$$+$$f$-space. This is obtained from the following 6$\times$6 matrix
\begin{equation} \label{eq:G0knu}
G^0(\boldsymbol{k},i\omega) \! = \!\!
\begin{pmatrix} 
i\omega \mathbb{I}_{2\times2} \! -\! H^{ff} & - H^{fc}(\boldsymbol{k})\\
\\
-H^{cf}(\boldsymbol{k}) & i\omega 
\mathbb{I}_{4\times4}  \!-\!H^{cc}(\boldsymbol{k}) 
\end{pmatrix}^{\hspace{-0.25cm}-\!1}.
\end{equation}
via the sum over $\boldsymbol{k}$:
\begin{equation} \label{eq:Gloc}
G^0_\text{loc}(i\omega) =\sum_{\boldsymbol{k}} G^0(\boldsymbol{k},i\omega).
\end{equation}
Using standard expressions for the block-matrix inversion, we extract the 2$\times$2 $ff$-block out of the 6$\times$6 local Green's function in Eq.~\ref{eq:Gloc}:  
\begin{equation} 
\left. G^0_\text{loc}(i\omega) \right|_{ff}=
\label{eq:Gloc_ff}
\sum_{\boldsymbol{k}} \frac{1}{i\omega \mathbb{I}_{2\times2}  \!-\! H^{ff} \!-\! H^{fc}(\boldsymbol{k}) \frac{1}{i\omega \mathbb{I}_{4\times4} -H^{cc}(\boldsymbol{k})} H^{cf}(\boldsymbol{k})},
\end{equation}
which will then have to be inverted and plugged into Eq.~\ref{eq:hyb_fun} to eventually arrive at $\Delta^{ff}(i\omega)$.

At the level of the DMFT impurity solver, it is important to know whether or not the hybridization function is diagonal within the $ff$-subspace. In order to assess the role of the strain and relaxation terms in determining the matrix structure of $\Delta^{ff}(i\omega)$, one convenient way is to perform a large-$\omega$ expansion of Eq.~\ref{eq:Gloc_ff}.  To do so, we note that the expression inside the sum over $\boldsymbol{k}$ in Eq.~\ref{eq:Gloc_ff} is the non-interacting Green's function of the following impurity model
\begin{equation} \label{eq:H_AIM}
\mathcal{H}^\text{AIM}  =  H^{ff}f^\dagger f + \sum_\alpha \left[ H^{cf}_\alpha(\boldsymbol{k}) f^\dagger c_\alpha(\boldsymbol{k}) + H^{fc}_\alpha(\boldsymbol{k}) c^\dagger_\alpha(\boldsymbol{k}) f \right] + 
\sum_{\alpha \beta} H^{cc}_{\alpha \beta}(\boldsymbol{k}) c^\dagger_{\alpha}(\boldsymbol{k})c(\boldsymbol{k})_{\beta}^{\phantom{dagger}} \,.
\end{equation}
In $\mathcal{H}^\text{AIM} $ the spin degrees have been neglected for simplicity, the indices $\alpha$ and $\beta$ run from 1 to 2, and the matrix elements have been written as 2$\times$2 blocks of the corresponding matrices appearing in Eq.~\ref{eq:Hfull}. Thus, for instance, the ${2\times 4}$ rectangular matrix reads
\begin{equation}\label{eq:Hfc_block}
 H^{fc}(\boldsymbol{k}) = \left( H^{fc}_{\alpha=1}(\boldsymbol{k}) , H^{fc}_{\alpha=2}(\boldsymbol{k})  \right).
\end{equation}
With the Hamiltonian \ref{eq:H_AIM}, we can easily calculate the moment expansion of the corresponding non-interacting Green's function via the commutators of $\mathcal{H}^\text{AIM}$ with $f$ and then evaluate order by order in 1/$\omega$ the anticommutator with $f^\dagger$ (see, for instance, Ref.~\cite{wagnerMottInsulatorsBoundary2023}). This way, we can perform the sum over $\boldsymbol{k}$ in Eq.~\ref{eq:Gloc_ff} after the $1/\omega$-expansion and obtain 
\begin{equation}\label{eq:Gloc_ff_exp}
\left. G^0_\text{loc}(i\omega) \right|_{ff}\! =\!
\frac{1}{i\omega}\! \left[ \mathbb{I}_{2\times 2} + 
\frac{C_1}{i\omega} -  
\frac{C_2}{(i\omega)^2} + \frac{C_3}{(i\omega)^3}  + ...
\right],
\end{equation}
where the $C_{n=1,2,3,...}$ have been defined as $(-1)^{n+1}$ times the sum over $\boldsymbol{k}$ of the coefficients at the $1/({i\omega})^n$-order expansion in Eq.~\ref{eq:Gloc_ff_exp}:
\begin{equation}\label{eq:C1}
C_1 = H^{ff}
\end{equation}
\begin{equation}\label{eq:C2}
C_2 = - \sum_{\boldsymbol{k}}  \left[  \left( H^{ff} \right)^2 + \sum_\alpha H^{fc}_\alpha(\boldsymbol{k}) H^{cf}_\alpha(\boldsymbol{k}) \right]
\end{equation}
\begin{equation}\label{eq:C3}
C_3 = \sum_{\boldsymbol{k}} \left[ \left(H^{ff} \right)^3  +  H^{ff}\sum_\alpha H^{fc}_\alpha(\boldsymbol{k}) H^{cf}_\alpha(\boldsymbol{k})  + \sum_\alpha H^{fc}_\alpha(\boldsymbol{k}) H^{cf}_\alpha(\boldsymbol{k}) H^{ff}+
\sum_{\alpha \beta}H^{fc}_\alpha(\boldsymbol{k}) H^{cc}_{\alpha \beta}(\boldsymbol{k}) H^{cf}_\beta(\boldsymbol{k}) \right],
\end{equation}
in which $H^{ff}$, being independent of $\boldsymbol{k}$, could be taken out of the sum.
We can then invert Eq.~\ref{eq:Gloc_ff_exp} keeping all terms at the corresponding order in $1/({i\omega})$:
\begin{align}\label{eq:invGloc_ff_exp}
\Big[ \left. G^0_\text{loc}(i\omega) \right|_{ff} \Big]^{-1} = i\omega \left[ \mathbb{I}_{2\times 2}  - \frac{C_1}{i\omega} + \frac{C_2 + C_1^2}{(i\omega)^2} - \frac{C_3 + C_1 C_2 + C_2 C_1 + C_1^3}{(i\omega)^3} + \mathcal{O}\left(\frac{1}{i\omega}\right)^4 \right].
\end{align}
This way, we arrive at the final expression for the $1/\omega$-expansion of     Eq.~\ref{eq:hyb_fun}:
\begin{align}\label{eq:hyb_fun_exp}
\Delta^{ff}(i\omega) &= \frac{1}{i\omega} \left(
\sum_{\boldsymbol{k}} \Big[  \left( H^{ff} \right)^2 + \sum_\alpha H^{fc}_\alpha(\boldsymbol{k}) H^{cf}_\alpha(\boldsymbol{k}) \Big] - \Big[ \sum_{\boldsymbol{k}}  H^{ff} \Big]^2  \right) +\frac{1}{(i\omega)^2} \left(  \sum_{\boldsymbol{k}} \Big[ \left( H^{ff}\right)^3 \right. + \nonumber\\
& + \big(H^{ff} \big)\! \cdot\! \big(\sum_\alpha H^{fc}_\alpha(\boldsymbol{k}) H^{cf}_\alpha(\boldsymbol{k})\big)  + 
\text{same with opposite mult.~order}+
\sum_{\alpha \beta}H^{fc}_\alpha(\boldsymbol{k}) H^{cc}_{\alpha \beta}(\boldsymbol{k}) H^{cf}_\beta(\boldsymbol{k}) \Big] +   \nonumber\\
&  - \Big(\sum_{\boldsymbol{k}} H^{ff}\Big) \! \cdot\! \Big(  \sum_{\boldsymbol{k}} \Big[  \left( H^{ff} \right)^2 + \sum_\alpha H^{fc}_\alpha(\boldsymbol{k}) H^{cf}_\alpha(\boldsymbol{k}) \Big] \Big)  -\text{same with opposite mult.~order} \, +\nonumber\\
& \left. + \Big[ \sum_{\boldsymbol{k}}   H^{ff} \Big]^3 \right) +\mathcal{O}\left(\frac{1}{i\omega}\right)^3,
\end{align}
where, again, since $H^{ff}$ has either zero or at most constant-in-$\boldsymbol{k}$ matrix elements, there are several cancellations. For the same reason, $H^{ff}_\text{loc}$ in the definition of $\Delta^{ff}$ (Eq.~\ref{eq:hyb_fun}) compensates $H^{ff}$ appearing in $C_1$ (see Eqs.~\ref{eq:C1} and \ref{eq:invGloc_ff_exp}).
Let us first review the THF starting point with no strain and no relaxation. 
Since $H^{ff}$ vanishes, the $1/i\omega$-term in Eq.~\ref{eq:hyb_fun_exp} has contributions only from the scalar product of the 2$\times$4 row vector $H^{fc}$ with the 4$\times$2 column vector $H^{cf}$. These amount to
\begin{equation}\label{eq:1st_nostrain}
\left(  \gamma^2 + {v^\prime}^2 \sum_{\boldsymbol{k}} (k_x^2+ k_y^2) \right) \mathbb{I}_{2\times2}
\end{equation}
The coefficient of the second-order term vanishes, because all terms containing $H^{ff}$ trivially drop out and the remaining one is the scalar product between two orthogonal vectors: 
\begin{equation}\label{eq:v1}
H^{fc}_\alpha(\boldsymbol{k}) = \Big(
    \gamma \sigma_0 \!+\! v^\prime \boldsymbol{k}\!\cdot\! \boldsymbol{\sigma} ,\,  0_{2\times 2} \Big)
\end{equation}
 and 
 \begin{equation}\label{eq:v2}
 \sum_\beta  H^{cc}_{\alpha \beta}(\boldsymbol{k}) H^{cf}_\beta(\boldsymbol{k})\! =\!\!
\left( \!\!\!
\begin{array}{c}
0_{2\times 2} \\
 \\ 
 v \boldsymbol{k}\!\cdot\!(\sigma_0, \!-i\sigma_3)\!
\left( \gamma \sigma_0 \!+\! v^\prime \boldsymbol{k}\!\cdot\! \boldsymbol{\sigma} \right)
\end{array}
 \! \! \!\right)\!\!.
 \end{equation}

\noindent

When strain is introduced, $\Delta^{ff}(i\omega)$ is not diagonal anymore. 
Adding Eq.~\ref{app:eqn:H_strain} to our 6$\times$6 $f$+$c$ Hamiltonian, we get individual contributions to $C_1$ from the $ff$-block which nevertheless cancel each other due to their momentum independence:
\begin{equation}
\sum_{\boldsymbol{k}} \left( H^{ff}  (\boldsymbol{k}) \right)^2 = \Big[
\sum_{\boldsymbol{k}} H^{ff}  (\boldsymbol{k})
\Big]^2 = M_f^2 \epsilon_-^2
\end{equation}
The remaining term at order $1/i\omega$ in Eq.~\ref{eq:hyb_fun_exp} contribute two -- still diagonal -- terms, one proportional to ${\gamma'}^2$ and one proportional to $\epsilon_-^2$, in addition to Eq.~\ref{eq:1st_nostrain}:
\begin{equation}\label{eq:strain_contr}
\sum_\alpha H^{fc}_\alpha(\boldsymbol{k}) H^{cf}_\alpha(\boldsymbol{k}) = \left(  \gamma^2 + {v^\prime}^2 \sum_{\boldsymbol{k}} (k_x^2+ k_y^2)  + {\gamma'}^2 \epsilon_+^2  + \left( c{''} \epsilon_-  \right)^2
\right) \mathbb{I}_{2\times2}.
\end{equation}
The crucial modifications due to strain, namely the $ff$ off-diagonal terms, manifest at the second order in $1/i\omega$.
$(H^{ff})^3$ is now finite but it is given by a constant times $\sigma_2$, and therefore there are still cancellations removing most of the contributions to the $(1/i\omega)^2$-term in Eq.~\ref{eq:hyb_fun}, with the exception of the following one:
\begin{equation} \label{eq:2nd_order_term}
\sum_{\alpha \beta}H^{fc}_\alpha(\boldsymbol{k}) H^{cc}_{\alpha \beta}(\boldsymbol{k}) H^{cf}_\beta(\boldsymbol{k}) \sim \Big( \gamma^2 c \epsilon_-  \Big) \sigma_2
\end{equation}
plus corrections of order $\epsilon^2$ with both $\sigma_1$ and $\sigma_2$ structure.
This represents a process in which an electron from the $f$-orbital hops onto the $c$ subspace, visits the dispersive $c$-bands and hops back onto the $f$. It is enabled by strain (compare to Eq.~\ref{eq:v1} and ~\ref{eq:v2} for the case without strain) and it appears in the 2$^{\rm nd}$-order term of the hybridization function (\ref{eq:hyb_fun}).

We now show that a similar conclusion can be alternatively reached using symmetry arguments in combination with a perturbative expansion in the strength of the strain. The most general strain term in the THF model contains the isotropic strain parameter $\epsilon_{+}$, the anisotropic $\epsilon_{-}$ and the shear $\epsilon_{xy}$. These transform as $x^2+y^2$, $x^2-y^2$ and $xy$ respectively under spatial symmetry. $f$ electron forms the $\Gamma_3$ irreducible representation~\cite{songMagicAngleTwistedBilayer2022,herzog-arbeitmanHeavyFermionsEfficient2024}. 
We expand the local hybridization function in powers of the strength of strain, and discuss the symmetry properties of each term.  
Without loss of generality, we focus on valley $\eta = +$ and spin $s$. 
The local hybridization can be written as 
\ba 
S_{hyb}^{\eta=+, s} = \sum_{n,m} \sum_{i\omega,\RR,\alpha\beta,\mu} f^{\dagger}_{\RR, \alpha, +, s }(i\omega) f_{\RR,\beta,+, s}(i\omega) 
\Delta^{\mu, n,m,i}_{+ s}(i\omega) \mathcal{C}_{\mu}^{n,m,i} (\epsilon_+)^n
(\epsilon_-)^m (\epsilon_{xy})^i [\sigma^{\mu}]_{\alpha\beta}
\ea 
where $\mathcal{C}_\mu^{n,m,i}$ denotes the coefficient of the contribution at the order of $ (\epsilon_+)^n
(\epsilon_-)^m (\epsilon_{xy})^{i}$ with Pauli matrix $\sigma^\mu$. $\Delta_{+ s}^{\mu,n,m,i}(i\omega)$ denotes the corresponding hybridization function. 
The symmetry-allowed terms up to $i+n+m\le 2$ are the following:
\begin{itemize}
    \item For $n=m=i=0$, the only allowed term is the $\sigma^0$ which represents the original contribution: 
    \ba 
    \sum_{i\omega,\RR,\alpha} f^{\dagger}_{\RR, \alpha,+, s }(i\omega) f_{\RR,\alpha,+, s}(i\omega) 
\Delta^{(0)}_{+ s}(i\omega)
\label{eq:strain_hyb_0}
    \ea 
    where we use $\Delta^{(0)}_{+ s}(i\omega)$ to denotes its zero-th order contribution. 
    \item For $n+m+i=1$, the allowed terms are proportional to $\epsilon_+ \sigma_0, \epsilon_{-} \sigma_2 + \epsilon_{xy}\sigma_1$ and can be written as 
    \ba 
     \sum_{i\omega,\RR,\alpha} f^{\dagger}_{\RR, \alpha, +, s }(i\omega) f_{\RR,\alpha,+, s}(i\omega) \epsilon_+
\Delta^{(1),1}_{+ s}(i\omega) 
+  \sum_{i\omega,\RR,\alpha,\beta} f^{\dagger}_{\RR, \alpha, +, s }(i\omega) f_{\RR,\alpha,+, s}(i\omega) \bigg( 
\epsilon_{-} \sigma_2 + \epsilon_{xy}\sigma_1\bigg)_{\alpha,\beta}
\Delta^{(1),2}_{+ s}(i\omega)\, .
\label{eq:strain_hyb_1}
    \ea 
    where $\Delta^{(1),1}_{+ s}(i\omega), \Delta^{(1),2}_{+ s}(i\omega)$ to denotes the first-order (in strain) contributions.  
    \item For $n+m+i=2$, the allowed terms are proportional to $\epsilon_+^2 \sigma_0 , \epsilon_+(\epsilon_{-} \sigma_2 + \epsilon_{xy}\sigma_1), (\epsilon_{-}^2+\epsilon_{xy}^2)\sigma_0, (\epsilon_{xy}^2-\epsilon_{-}^2)\sigma_2 + (-2\epsilon_{xy}\epsilon_-)\sigma_1 $, where we note that $\bigg((\epsilon_{xy}^2-\epsilon_{-}^2),(-2\epsilon_{xy}\epsilon_-)\bigg)$ form the same irreducible representation as $(\epsilon_{xy},\epsilon_{-})$.  The generic hybridization function can be written as 
    \ba 
     \sum_{i\omega,\RR,\alpha\beta} f^{\dagger}_{\RR, \alpha, +, s }(i\omega) f_{\RR,\beta,+, s}(i\omega)&
     \bigg[ 
     \epsilon_+^2 \sigma_0 \Delta_{+s}^{(2),1}
     +\epsilon_+(\epsilon_-\sigma_2+\epsilon_{xy}\sigma_1)\Delta_{+s}^{(2),2} 
     +(\epsilon_-^2+\epsilon_{xy}^2)\sigma_0\Delta_{+s}^{(2),3}  \nonumber\\ 
     &
       +\bigg( (\epsilon_{xy}^2-\epsilon_{-}^2)\sigma_1 
       +(-2\epsilon_{xy}\epsilon_-)\sigma_2 \bigg) 
       \Delta_{+s}^{(2),4}  
     \bigg]_{\alpha\beta}
     \label{eq:strain_hyb_2}
    \ea 
     where $\Delta^{(2),1}_{+ s}(i\omega), \Delta^{(2),2}_{+ s}(i\omega),\Delta^{(2),3}_{+ s}(i\omega), \Delta^{(2),4}_{+ s}(i\omega)$ to denotes the second-order (in strain) contributions.  
\end{itemize}
For the current situation, we focus on the limit where $\epsilon_{xy}=0$. Therefore, up to second order in the strength of strain, we have the following terms (from \cref{eq:strain_hyb_0,eq:strain_hyb_1,eq:strain_hyb_2})
\ba 
\sum_{i\omega,\RR,\alpha\beta} f^{\dagger}_{\RR, \alpha, +, s }(i\omega) f_{\RR,\beta,+, s}(i\omega)&
     \bigg[ 
     \sigma_0  \Delta^{(0)}_{+ s}(i\omega) 
     + \epsilon_-\sigma_2 \Delta^{(1),2}_{+s}(i\omega) \nonumber\\ 
     &
     + \epsilon_+^2 \sigma_0 \Delta^{(2),1}_{+s}(i\omega) 
     +\epsilon_+\epsilon_- \sigma_2  \Delta^{(2),2}_{+s}(i\omega) 
     + \epsilon_-^2 \sigma_0 \Delta^{(2),3}_{+s}(i\omega) 
     - \epsilon_-^2 \sigma_1 \Delta^{(2),4}_{+s}(i\omega) 
     \bigg]_{\alpha\beta}
\ea 
The final structures of the Pauli matrices at each order in $\epsilon$ are consistent with the results obtained from the $1/(i\omega)$ expansions.

\section{Basis rotation for DMFT calculation}\label{sec:DMFTbasis}

For the numerical solution of the THF model we employ the \textit{w2dynamics} software suite~\cite{wallerbergerW2dynamicsLocalOne2019}, which belongs to the class of DMFT-CTQMC solvers. While extremely robust and able to work with continuous baths and in an extended temperature range, this class of solvers famously suffers from a limitation called fermionic sign problem (see, e.g. Ref.~\cite{gullRMP}). 
No exact prescription for the improvement of the quality of the Monte Carlo sign exists, but it is generally true that off-diagonal local impurity Hamiltonians and hybridization functions tend to aggravate the problem \cite{shinaokaPRB}.
In addition to a generally better Monte Carlo sign, another advantage of working with orbital/valley/spin-diagonal impurity problems is the necessity of evaluating a smaller number of matrix elements for the local Green's function and self-energy. Furthermore, upon postprocessing there is no need to analytically continue the off-diagonal elements of $G$ and $\Sigma$, which is a particularly delicate process.

In this respect, the unstrained and unrelaxed THF model is particularly well-suited for CTQMC simulations, as its impurity Hamiltonian is identically zero, apart from effective chemical potential terms (cfr Sec.~\ref{sec:DMFTsolution}), and the hybridization function is diagonal in the orbital, valley and spin degrees of freedom, yielding a diagonal Green’s function for the impurity problem in the absence of spontaneous symmetry breaking.
As explained in Sec.~\ref{sec:HybFunc}, this is no longer the case in presence of strain and relaxation, which, in the original THF basis, cause the insurgence of off-diagonal terms in  both the impurity Hamiltonian ($H^{ff}$) and $\Delta^{ff}(\omega)$.

To briefly recap, the effect of relaxation terms is to introduce an on-site energy shift for the $\Gamma_{1} + \Gamma_{2}$ and $\Gamma_{3}$ $c$-orbitals of value $\mu_{1}=14.4meV$ and $\mu_2=4.5 meV$.
The effect of strain on the $f$-electrons is to add intra-$f$ orbital hybridization with the form
\begin{equation} \label{eq:deltaHstrain}
    \delta H^{\eta}_{\epsilon}= M_{f}\epsilon_{-}\sigma_{2}(-1)^{\eta}
\end{equation}
as well as additional $f-c$ hybridization terms, proportional to $\sigma_{3}$. 

Due to the symmetries of the strained and relaxed THF model, there is no basis in which $\Delta^{ff}(\omega)$ can be made diagonal at all frequency, including the basis that diagonalizes $H^{ff}$.
Nevertheless, this turns out to be a good approximation for our problem, as we will show in the following.  
The corresponding rotation matrix is
\begin{equation}
    U^{\eta}=\begin{bmatrix}
(-1)^{\eta+1}i\frac{\sqrt{2}}{2} & (-1)^{\eta}i\frac{\sqrt{2}}{2}  \\
\frac{\sqrt{2}}{2}  & \frac{\sqrt{2}}{2} 
\end{bmatrix}_{ff}\oplus \mathbb{I}^{\eta}_{cc}
\label{eq:rotation_matrix}
\end{equation}
for $\eta=\pm1$ valley index. This gives by construction an orbital-diagonal local problem but, in light of the analysis of Sec.~\ref{sec:HybFunc}, it turns out to diagonalize also the biggest off-diagonal contribution of $\Delta^{ff}(\omega)$. 
The reason for this is ultimately that on the one hand $\delta H^{\eta}_{\epsilon}$ entering the local problem with strain (see Eq.~\ref{eq:deltaHstrain}) and on the other hand the dominant off-diagonal terms in $\Delta^{ff}(\omega)$ are both proportional to $\sigma_2$ (see Eq.~\ref{eq:2nd_order_term}). The latter are hence exactly rotated away by the transformation (\ref{eq:rotation_matrix}). 
Note that there are terms in $\Delta^{ff}(\omega)$ at the same but also at lower order $1/\nu$ in the Matsubara frequency expansion (see previous section), that were diagonal in the original basis but become now off-diagonal after the basis transformation.
Yet, as we show below, there exists a range of parameters, namely strain percentage and temperature, where such terms do not grow beyond a given threshold. 
To verify this, we will analyze the dependence on the strain parameter $\epsilon$ and temperature $T$ of the hybridization function, comparing three characteristic scales:

\begin{itemize}
\item Local $f$-band energy splitting in the rotated basis
\item Maximum modulus of the diagonal component of the hybridization function
\item Maximum modulus of the off-diagonal component of the hybridization function
\end{itemize}

In particular, the last one has to be compared with the first and second, since it determines the deviation from a purely diagonal form of the Green's function for the $f$ subspace, as detailed in section~\ref{sec:HybFunc} with particular reference to Eq.~\ref{eq:Gloc_ff}.


\subsection{Effect of basis rotation on the local impurity Hamiltonian}

\begin{figure}[h]
  \centering
  \includegraphics[width=0.4\linewidth]{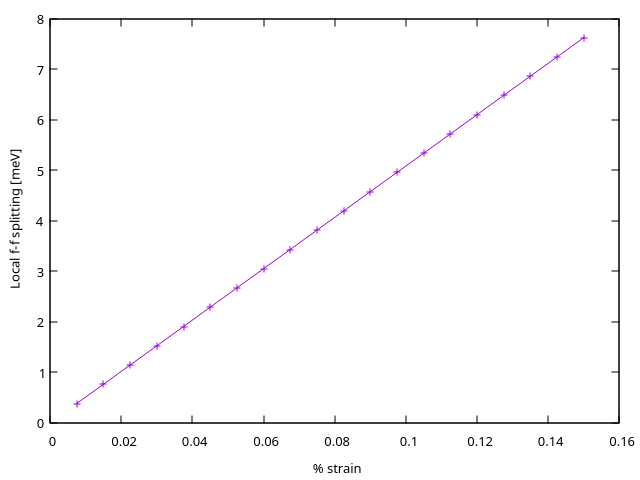}
  \caption{Local $f$-band splitting as a function of strain}
  \label{fig:local_splitting}
\end{figure}

In the rotated basis, the $f$-subspace consists of doubly-degenerate levels, whose local splitting is $2 M_{f}\epsilon_{-}$. This splitting is constant in temperature and linear in strain, as described in~\cite{herzog-arbeitmanHeavyFermionsEfficient2024} and visualized in Fig.~\ref{fig:local_splitting}. 
 
\subsection{Effect of basis rotation on the hybridization function}

The hybridization function on the Matsubara axis, which is the quantity entering the Monte Carlo simulations together with the impurity $H$, is given by ~\ref{eq:hyb_fun}.
In the rotated basis, this quantity is in general complex and has has a diagonal and an off-diagonal part. The off-diagonal components are all equal in modulus, hence we can take only one representative of these terms to compare with two non-degenerate diagonal components. Our goal is to perform CTQMC simulations discarding the off-diagonal components of $\Delta$, so their relative size with respect to the diagonal ones reflects the accuracy of our approximation.
In the following we study the behavior of the diagonal and off-diagonal components of the hybridization function depending on strain and temperature.

\begin{figure}[h]
  \centering
  \subfloat[]{
    \includegraphics[width=0.48\linewidth]{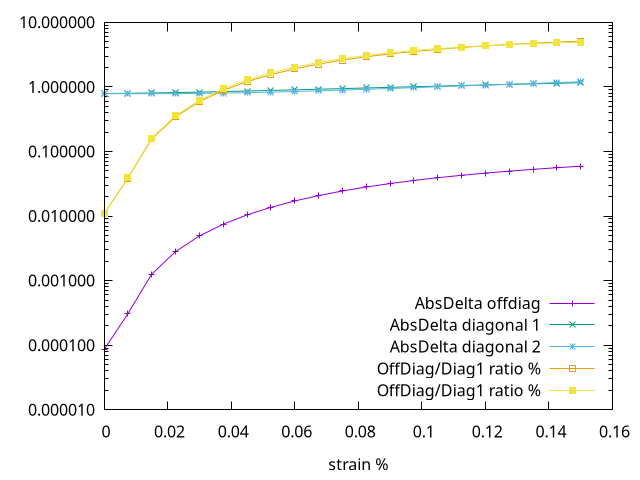}
  }
  \hfill
  \subfloat[]{
    \includegraphics[width=0.48\linewidth]{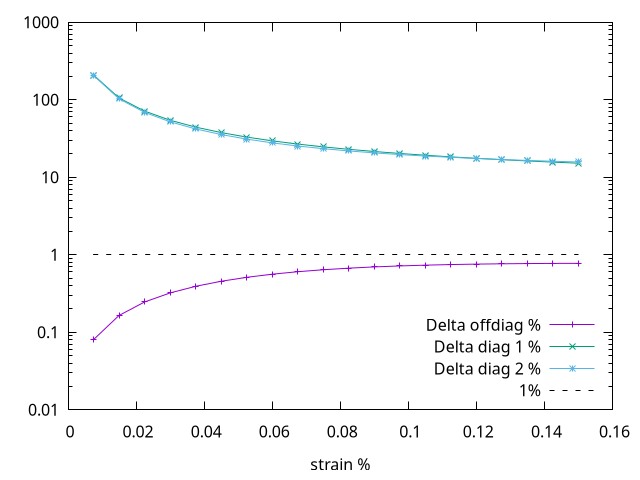}
  }
  \caption{(a) Absolute and relative magnitude of the diagonal and off-diagonal components of the hybridization function evaluated at the first Matsubara frequency as a function of strain (b) Diagonal and off-diagonal components of Delta  at the first Matsubara frequency as a percentage of the local $f$-band energy splitting, as a function of the strain percentage.}
  \label{fig:delta_vs_splitting}
\end{figure}

\subsubsection{Strain dependence}

Since at the chosen temperature $T=11.6 K$ the hybridization function is maximum at the first Matsubara frequency, we compare the relative magnitude (taken as the absolute value $|\Delta|=\sqrt{\mathrm{Re}\Delta^{2} + \mathrm{Im}\Delta^{2}}$) of the diagonal and off-diagonal components at this frequency. In Fig.~\ref{fig:delta_vs_splitting}(a) we plot the absolute value of the diagonal and off-diagonal components of $\Delta$ at the first Matsubara frequency as a function of the strain parameter. We also plot the ratio between them as a percentage. 

In Fig.~\ref{fig:delta_vs_splitting}(b) we instead show the ratio between the diagonal and off-diagonal components of $\Delta$ and the local $f$-band energy splitting as a function of strain. The data are represented as a percentage of the $f$-band energy splitting.

From these comparisons, it is clear that the off-diagonal component of $\Delta$ represents at most a perturbation of the order of $10\%$ of the diagonal component, and never increases above $1\%$ of the local $f$-level energy splitting. 

\subsubsection{Temperature dependence}

\begin{figure}[h]
  \includegraphics[width=0.48\linewidth]{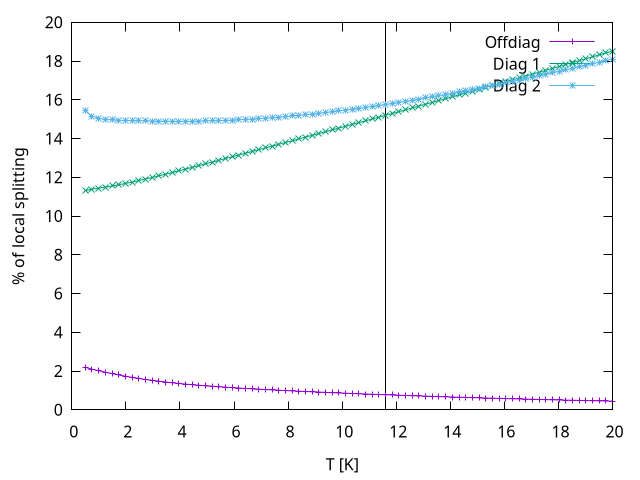}
  \caption{Modulus of the diagonal and off-diagonal components of $\Delta(i\omega_{1})$ as a function of temperature. }
  \label{fig:delta_reim}
\end{figure}

We now fix the strain parameter at the value $\epsilon=0.15\%$ chosen in the main text and consider the effect of temperature on the relative magnitude of the diagonal and off-diagonal components of $\Delta$. The results are plotted in Fig.~\ref{fig:delta_reim}, where the absolute values of the diagonal and off-diagonal $\Delta$ components at the first Matsubara frequency are plotted as a percentage of the local $f$-band energy splitting and as a function of $T$. It is immediate to notice how, for a strain of $0.15\%$, the off-diagonal component remains a rather small perturbation with respect to the dominant impurity energy scale, increasing to around $2\%$ for values of $T$ smaller than $0.5K$, way lower than the optimal temperature range for CTQMC and inside the ordered phase, not considered in this work. By contrast, the approximation of a diagonal $\Delta$ becomes better and better with increasing $T$, since thermal broadening effects overcome the strain-induced $f$-manifold splitting, which is responsible for the off-diagonal hybridization terms. 


\subsection{Effect of basis rotation on the self-energy}

\begin{figure}[h]
  \centering
  \subfloat[]{
    \includegraphics[width=0.32\linewidth]{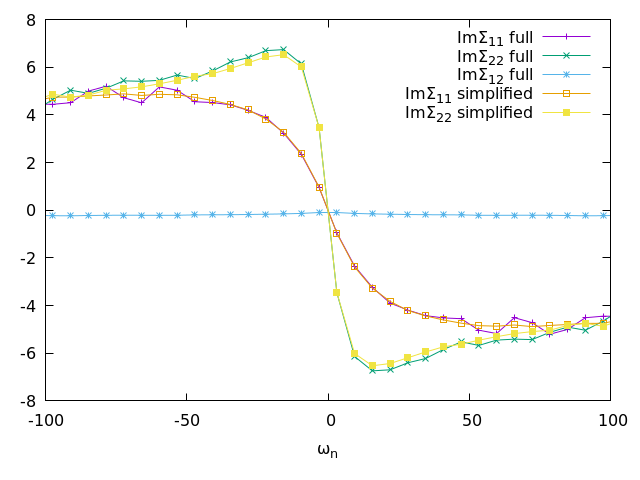}
  }
  \hfill
  \subfloat[]{
    \includegraphics[width=0.32\linewidth]{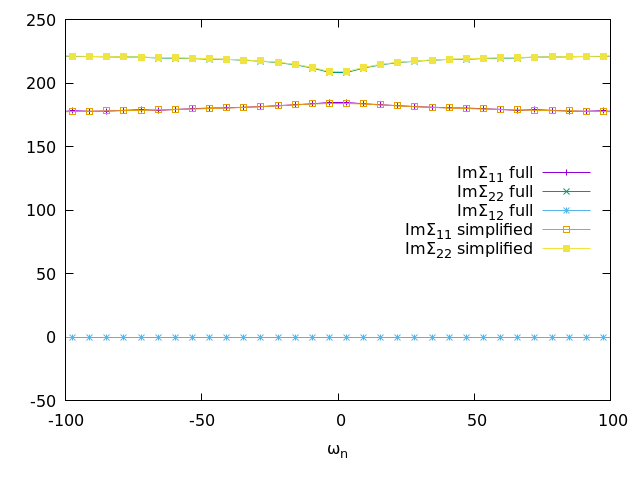}
  }
  \hfill
  \subfloat[]{
    \includegraphics[width=0.32\linewidth]{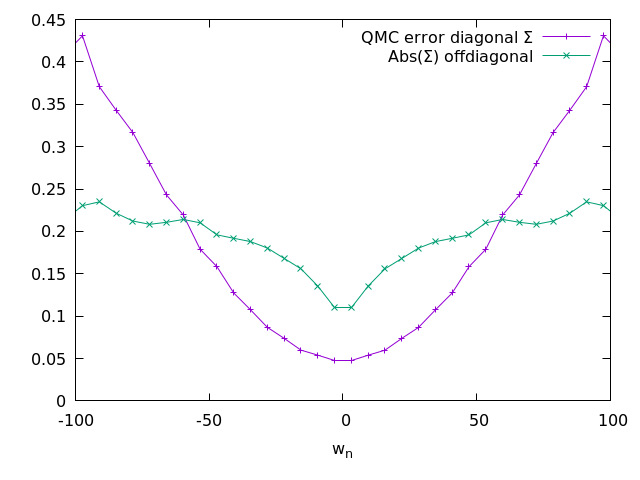}
  }
  \caption{(a) Imaginary part of $\Sigma(i\omega_{n})$, diagonal and off-diagonal components on the Matsubara axis. (b) Real part of $\Sigma(i\omega_{n})$, diagonal and off-diagonal components on the Matsubara axis. (c) Absolute value of the off-diagonal component of $\Sigma$ on the Matsubara axis for the ``full" simulation, compared with the QMC error on the diagonal component of $\Sigma$ for the first $f$ orbital ($\Sigma$ for the second orbital is of the same order of magnitude). }
  \label{fig:sigmas}
\end{figure}

While the hybridization function can give us a rough estimate of the degree to which the problem is off-diagonal, the physically relevant information comes from the local Green's function (or equivalently, the self-energy) of the problem. This is the product of the numerical simulation, and is obtained as the sum of a subset of scattering diagrams at all orders within the DMFT prescription. Only by assessing the relative strength of the diagonal and off-diagonal self-energy components can we have a precise estimation of the quality of our approximation.

In Fig.~\ref{fig:sigmas} we show CTQMC data for filling $\nu=-0.4$. The filling choice has no particular reason, and the behavior at all other fillings in the $[-4;4]$ range is analogous. 
We performed two sets of simulations in the rotated basis. In the first set, henceforth ``simplified", we apply the approximation of considering $\Delta(\i\omega_{n})$ diagonal at all frequencies. In the second (``full"), we run a fully off-diagonal simulation, capturing all components of $\Sigma$ on the Matsubara axis.

The first and second panel of Fig.~\ref{fig:sigmas} show a comparison between the real and imaginary parts of the self-energy. The following quantities are compared  (i) the orbital-diagonal $\Sigma$ components resulting from the ``full" simulations, for the two non-degenerate $f$-orbitals, (ii) one orbital off-diagonal components of $\Sigma$ resulting from the ``full" simulations (the others being related by symmetry), (iii) the orbital-diagonal $\Sigma$ components resulting from the ``simplified" simulations.

It is immediate to notice how, compared to the diagonal components, the off-diagonal components of $\Sigma$ represent just a minimal perturbation. Moreover, by comparing the diagonal components of the ``full" and ``simplified" simulations, it is clear that the approximation has only negligible effects on the diagonal self-energies.


In fact, the off-diagonal components of $\Sigma$ are comparable in magnitude with the QMC error on the diagonal component, which is the standard deviation of the results of the independent Monte Carlo simulations ($144$ in our case) as evidenced in panel (c).
We conclude that, especially for the values of strain and temperature at which our simulations are run, assuming a diagonal hybridization function is a justified approximation, and hence the basis rotation described in~\ref{eq:rotation_matrix} is effective in rendering the impurity problem diagonal for the purpose of DMFT simulations.

\section{Charge self-consistent DMFT solution}\label{sec:DMFTsolution}

We solve the strained and relaxed THF model within the framework of Dynamical Mean-Field theory~\cite{georgesDynamicalMeanfieldTheory1996}, making use of the Continuous-Time Quantum Monte Carlo (CTQMC) solver w2dynamics~\cite{wallerbergerW2dynamicsLocalOne2019}. 
As discussed in section~\ref{sec:InteractingH}, the intraction Hamiltonian is treated at two different levels of approximation: the $H_{U}$ is dynamically accounted for within the DMFT approximation, while the remaining ones are mean-field decoupled. Concordantly, the electronic self-energy of the lattice problem is obtained self-consistently: first, starting from the density matrix at each loop of the DMFT simulation, the mean-field terms are obtained. These are summed to the non-interacting Hamiltonian. The $f$-orbital peojection of this operator represents the impurity Hamiltonian which is solved via CTQMC. The resulting self-energy is upfolded in the combined $f\oplus c$ space by padding it with zeros, since no dynamical correlation effects are considered for the dispersive bands. Self-consistent update of the density matrix and self-energy are then performed until convergence. 
Since we treat all interaction terms apart from $H_{U}$ at the Hartree level, we can perform a further simplification~\cite{huSymmetricKondoLattice2023} that allows us to describe them as an effective ``double-counting" term, i.e. an orbital-selective chemical potential acting on the correlated subspace only:

\begin{equation}
   H_{W}+H_{V} = H_{DC} + H_{N}
\end{equation}

where

\begin{equation}
    H_{\mathrm{DC}}= \big(W (\nu_{f} - \nu_{c}) - V \nu_{c}\big)\big(\sum f^{\dagger}f\big) = \big(W (\nu_{f} - \nu_{c}) - V \nu_{c}\big)\hat{n}_{f} 
    \label{eq:H_DC}
\end{equation}

is responsible for the $f$-subspace energy shift. The second term is
\begin{equation}
    H_{\mathrm{N}} = \big(W \nu_{f} + V \nu_{c}\big)\big(\sum f^{\dagger}f + \sum c^{\dagger}c\big) = \big(W \nu_{f} + V \nu_{c}\big)\hat{N} 
    \label{eq:H_tot}
\end{equation}

 and it couples to the total occupation $\hat{N}$, having therefore the form of an effective chemical potential. We note that, since $W\approx V\approx 47meV$, this term is remarkably close to the expected energy contribution coming from the plane plate capacitor with inter-gate distance $\xi=10 nm$, which is 

\begin{equation}
e\Delta\Phi = \nu_{\mathrm{tot}} \frac{e^2\xi}{4\Omega_0 \epsilon_0 \epsilon_r} = \nu_{\mathrm{tot}}\cdot 47\,\rm meV.
\label{eq:geom_capa_SB}
\end{equation}

where $\Omega_{0}$ is the area of the mBZ and $\epsilon_{r}=6$.
Hence, by simply removing the term $H_{\mathrm{N}}$ from the interacting Hamiltonian we effectively remove the geometric capacitance from the simulation, hence offering a closer representation (e.g. in the compressibility and entropy calculations) of the quantum capacitance effects of the TBLG alone.

\subsection{Numerical simulation parameters}
We study the strained and relaxed THF model at $T=11.6 K$ ($\beta=1 meV$).
We perform DMFT simulations in the full doping range $\nu\in[-4;4]$, starting from the CNP, where the mean-field corrections are the weakest, and proceeding from previous converged solution in ascending and descending steps of $\Delta\nu=0.05$. 

The noninteracting Hamiltonian is rotated to the basis where the $k$-averaged $H^{ff}$ is diagonal, as detailed in~\ref{sec:DMFTbasis}. We perform a minimum of 50 DMFT steps at each point, assessing the achieved convergence by comparing the values of observables (e.g. occupations) at the latest 5 steps, and accepting a converged solution when variations are of the order of $10^{-4}$. Self-energy, chemical potential and effective double-counting correction are mixed between successive DMFT steps in a $1:1$ ratio.

\subsection{QMC parameters}

We run the simulations, consistently with~\cite{raiDynamicalCorrelationsOrder2024}, on 144 CPU cores, using ${\texttt{Nmeas}} = 7.5\cdot10^{4}$ measurements per core for each Monte Carlo simulation, with a number of steps ${\texttt{Ncorr}}=2000$ per core between successive measurements to avoid autocorrelation effects and ${\texttt{Nwarmup}}=2\cdot 10^{6}$ warmup steps per core before each simulation.
We measured in imaginary time domain and used the Legendre polynomial basis, with maximum order $\texttt{NLegMax}=40$.

\subsection{Analytic continuation}\label{app:continuation_w2dyn}
Analytical continuation is performed starting from the last DMFT run using the MaxEnt method from the \textit{ana\_cont} software suite~\cite{kaufmannAnaContPython2023}. Once the self-energy and Green's function are analytically continued, they can be rotated back in the original orbital basis, where $H^{ff}$ is not diagonal. 
We used a regular mesh of $1001$ real-axis frequency points in the interval $[-250:+250]$ meV for $A(\omega)$ and $[-300:+300]$ meV for $\Sigma(\omega)$ and no preblur.

\section{Momentum-resolved spectral function}

\begin{figure}
    \centering
    \includegraphics[width=\linewidth]{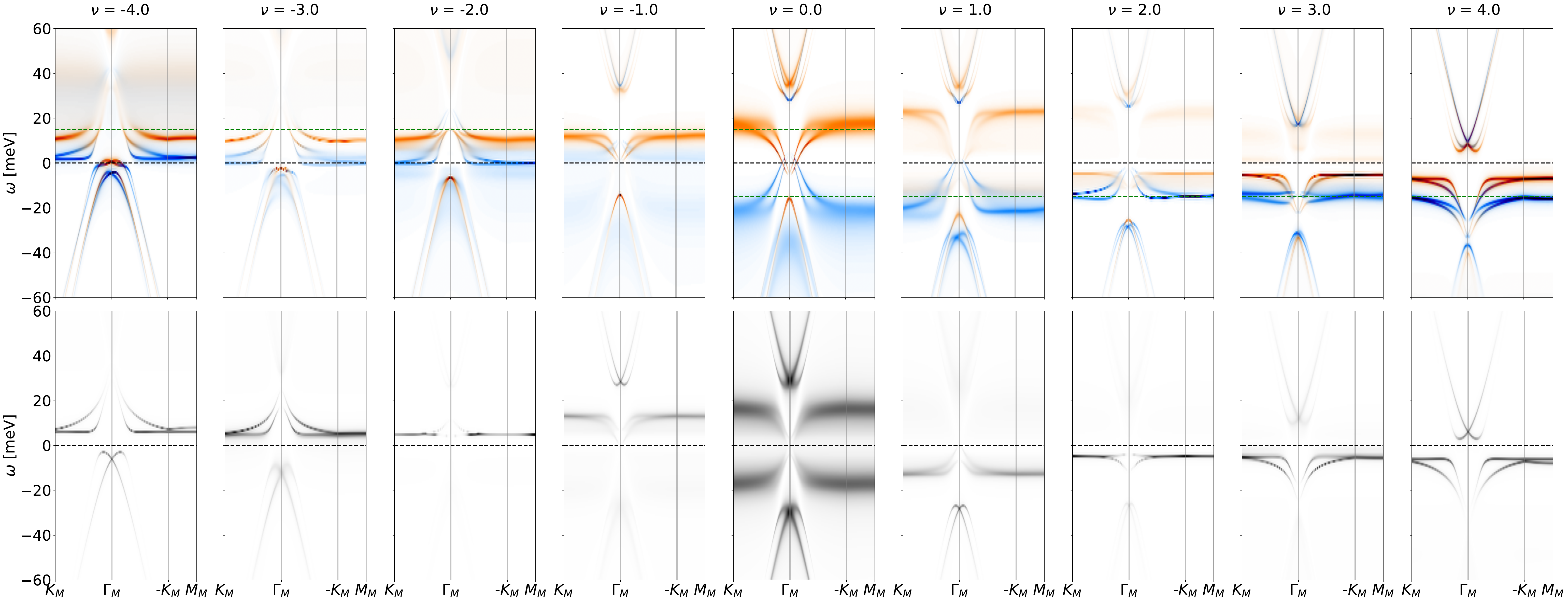}
    \caption{Momentum-resolved spectral function for the $f$ electrons at integer fillings. Second row: momentum-resolved spectral function for the $f$ electrons in absence of strain and relaxation.
    }
    \label{fig:akw-panels}
\end{figure}


\subsection{Spectral weight in $A(\mathbf{k},\omega)$ for integer fillings}

Fig. ~\ref{fig:akw-panels} shows the momentum-resolved spectral function along the $K_{M}\rightarrow\Gamma_{M}\rightarrow-K_{M}\rightarrow M_{M}$ high-symmetry path. The panels in the first row refer to all integer fillings in the range $\nu\in[-4;+4]$ 
The doping values have been chosen to be directly comparable to the experimental data in~\cite{xiaoInteractingEnergyBands2025}. The simulations are in a rather good agreement with the QTM results: they feature flat bands almost everywhere in the mBZ, apart from near the $\Gamma_{M}$ point where they become dispersive. At the CNP, the flat bands are separated by about 35 meV, in agreement with the experiment, and they touch near the $\Gamma_{M}$ point. At positive and negative integer $\nu$ values, a persistence spectral maximum due to the flat band is visible at negative and positive $\omega$ respectively, the value of which is in the range $\pm[10;20]$ meV. Increasing occupation from $\nu=-4$, the lower flat band first remains rather pinned at zero frequency up to $\nu=-1$, after which it shifts to lower frequency and becomes the negative-bias persistent feature for electron doping. 
Comparing positive and negative doping of the same size, it is interesting to notice how the electron-doped side features a larger spectral gap at large doping with respect to the hole-doped side: this is most evident for $\nu=+4$, where the system is in a clear band-insulating state. This is to be compared to the $\nu=-4$ case, which features a vanishing indirect gap at the Fermi level, and is a product of the $P$-symmetry breaking relaxation terms, which shift the $f$-bands asymmetrically within the hybridization gap. The $\nu=-4$ case is particularly delicate, since the presence or absence of the gap is directly linked to the model parameters, in particular the effective chemical potentials deriving from lattice relaxation.

\begin{figure}
    \centering
    \includegraphics[width=\linewidth]{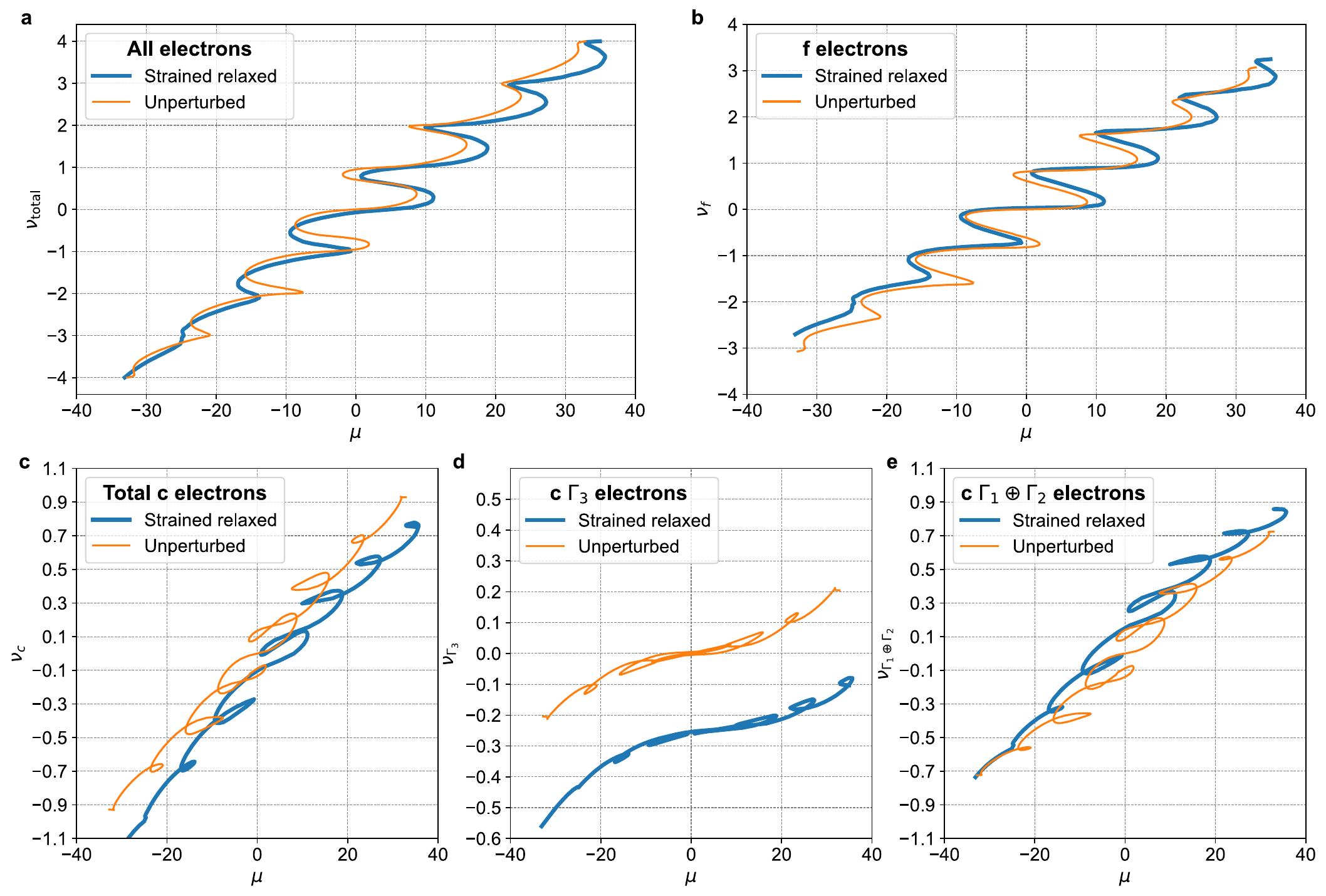}
    \caption{Occupation with respect to CNP as a function of the chemical potential, total value (a) and orbital-valley resolved (b-c), for T=11.6 K. The blue curves refer to DMFT simulations in presence of strain and lattice relaxation, while the orange one are the unperturbed results from~\cite{raiDynamicalCorrelationsOrder2024}.  }
    \label{fig:all_occupations}
\end{figure}

A further consequence of this asymmetric shift is evident from the comparison between the strained and unstrained bands at integer fillings (first and second row of Fig.\ref{fig:akw-panels}). Especially in the hole doping region, the combined effect of particle-hole symmetry breaking and f-c hybridization stabilizes a coherent $f$-electron spectral peak much nearer to the Fermi level for integer fillings than in the unperturbed case. This effect is most evident from $\nu = -4$ to $-2$. This doping range is precisely that for which the inverse compressibility peaks feature a marked reduction end depinning from integer filling. As noted in the main text, this is due to the increased metallic character of the solution, which is now evident from the spectral function analysis.
The increased metallicity of the system can also be evinced by comparing the $\mu(\nu)$ curves of the strained/relaxed and unperturbed systems, which is done in Fig.~\ref{fig:all_occupations}: as noted in the main text, for large negative dopings the system is less prone to exhibiding quantum dot-like behavios. Concordantly, the size of flat occupation plataux is greatly reduced with respect to the electron-doped case. An increase in $fc$ hybridization in the strained case is also evident from panels (b-d) of Fig.~\ref{fig:all_occupations}: when $\nu=-4$, i.e. when the flat bands are completely depleted, the $f$-occupation is noticeably larger than in the unperturbed case, accounting for an increase in $f$ character of the lower dispersive bands. For $\nu=+4$ the $f$-occupation is also higher: in this case, this reflects the enhanced gap between narrow and upper-dispersive bands.


\subsection{Spectral peaks in $A_{\Gamma_{M}}(\omega)$}

\begin{figure}
    \centering
    \includegraphics[width=0.5\linewidth]{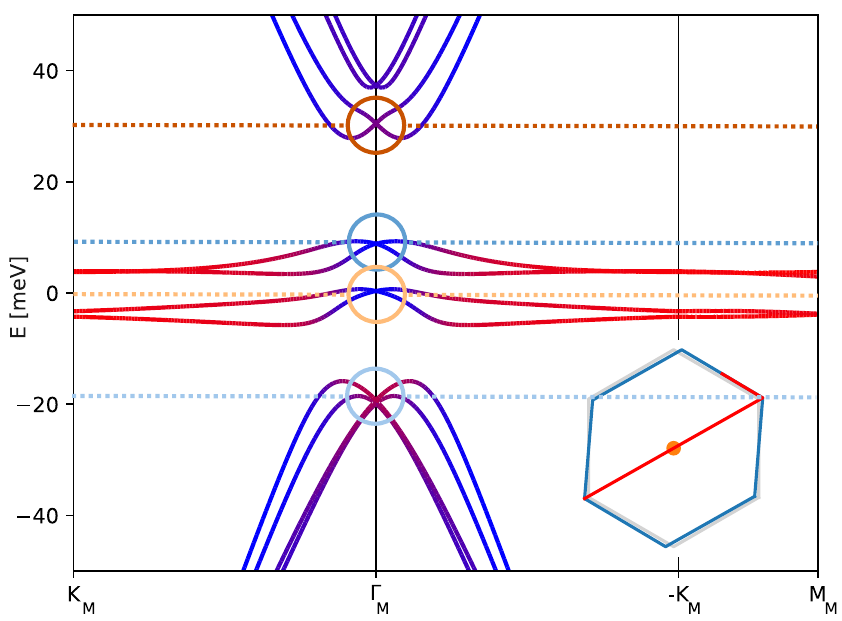}
    \caption{Noninteracting band structure showing the orbital character from $f$ (red) to $c$ (blue).
    }
    \label{fig:bands-orbital-character}
\end{figure}

\begin{figure}
    \centering
    \includegraphics[width=\linewidth]{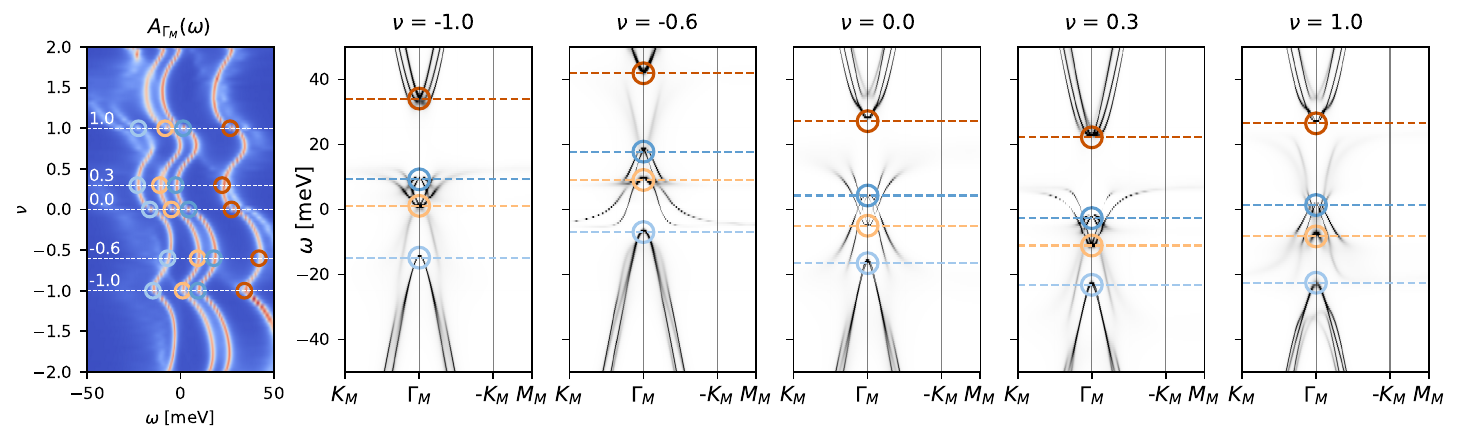}
    \caption{On the left, spectral function as a function of filling at the $\Gamma_{M}$. points. The 5 panels on the right are the momentum-resolved spectral functions of the $c$ electrons evaluated at the four fillings denoted by white lines in the left panel: $\nu = -1.0, -0.6, 0.0, 0.3, 1.0$. The circles highlight the positions of the spectral maxima.
    }
    \label{fig:aw_gamma_point}
\end{figure}

One of the innovations provided by the Quantum Twisting Microscope is the possibility of accessing the spectral function at different points in the Brillouin Zone. By probing $A(\omega)$ at the $\Gamma_{M}$ point, the authors of~\cite{xiaoInteractingEnergyBands2025} observe a spectral maximum inside the flat band gap, which evolves with doping in a manner reminiscent of the $\mu(\nu)$ curve (see Fig.1e) in the main text.

This behavior can be understood by observing the orbital character of these spectral peaks. In ~\ref{fig:bands-orbital-character} we plot the orbital character of the noninteracting bands for the THF model, in presence of strain and relaxation. Due to the topological obstruction featured by TBLG, the orbital character at the $\Gamma_{M}$ point is essentially $c$ (blue) in the narrow bands, while it is $f/c$ mixed at the top and bottom of the dispersive bands. The circles and dotted lines denote the positions and relative distance of the spectral weight at the $\Gamma_{M}$ point.

In Fig.~\ref{fig:aw_gamma_point}, in the left panel we show the spectral function at the $\Gamma_{M}$ point as a function of doping. This features four wiggling spectral maxima, that evolve from positive to negative frequency as a function of doping. The other panels of Fig.~\ref{fig:aw_gamma_point} show the momentum-resolved spectral functions for $c$-electrons at five different values of doping, including the CNP. The positions and relative distances of the peaks, are again represented by colored circles and dashed lines. It is immediate to notice that the circles all refer to local spectral maxima of $c$ character. By comparing the relative distances, it is also easy to notice that the four peaks are almost rigidly shifted upon varying the chemical potential.
This fact can be understood in the following way: since the $c$ electrons are essentially uncorrelated, their behavior of their spectral distribution is well approximated by that of noninteracting electronic bands. Varying the chemical potential by $\Delta\mu$ has the only effect of rigidly shifting the whole band structure by $-\Delta\mu$ with respect to zero frequency. This will follow the peculiar wiggling behavior that results from the $f/c$ spectral weight reshuffling, sometimes referred to as Dirac revivals, as described in~\cite{raiDynamicalCorrelationsOrder2024}.

In the noninteracting band diagram, the orange circle sits at zero energy, i.e. $µ_{CNP}$. Conconrdantly, the relative spectral peak in Fig.~\ref{fig:aw_gamma_point} precisely follows $-\mu(\nu)$ across the whole doping range. The other spectral peaks mirror this behavior with an energy offset of about $+35$ (red), $+8$ (blue) and $-15$ $meV$ (light blue) respectively. The experimental data from~\cite{xiaoInteractingEnergyBands2025} compare favorable with these results, showing a wiggling enhanced spectral weight region which follows the evolution of the chemical potential as a function of doping. The distance beetween the outmost peaks, $\approx 50 meV$, is correctly captured as well.

\section{Entropy calculations}

\begin{figure}
    \centering
    \includegraphics[width=0.5\linewidth]{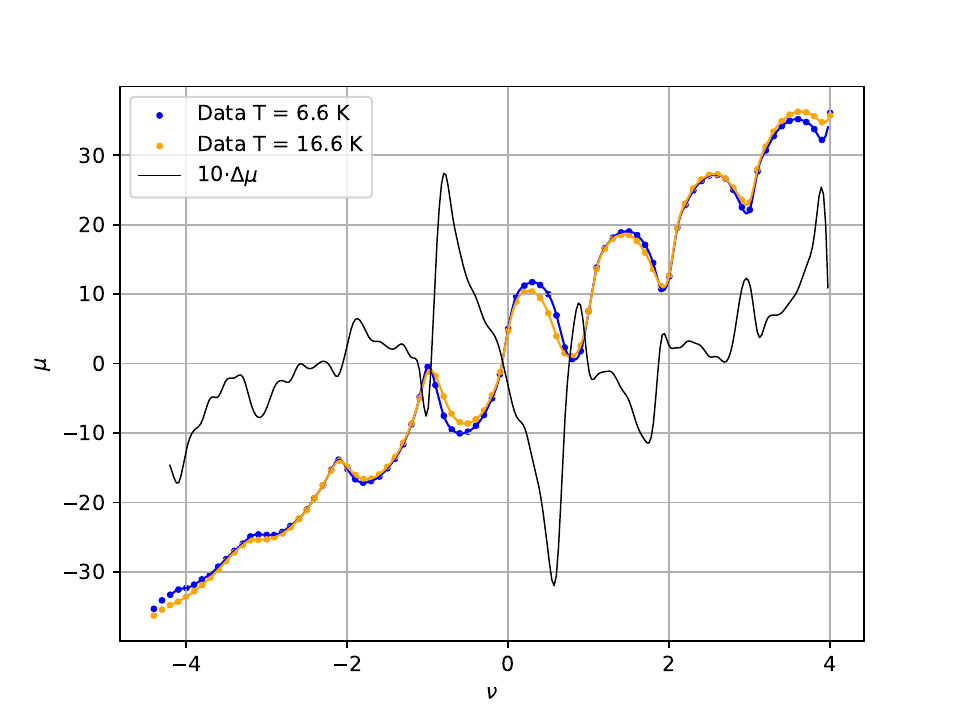}
    \caption{Chemical potential as a function of doping for the interval $\nu\in[-4,4]$.
    }
    \label{fig:mu_of_nu}
\end{figure}

Entropy is calculated by making use of the Maxwell relation
\begin{equation}
    S(\nu,T)=-\int_{4}^{\nu}\bigg(\dfrac{\partial \mu}{\partial T}\bigg)_{\nu'} \mathrm{d}\nu'
\end{equation}

To calculate $S(\nu,T)$, three DMFT-QMC simulations are made at inverse temperature $\beta = \frac{11.6}{T} [meV]$, $\frac{11.6}{T+\Delta T} [meV]$ and $\frac{11.6}{T-\Delta T} [meV]$ at all considered doping values $\nu$. The value of $\Delta T$ is empirically set at $5 K$, a value at which the chemical potential varies meaningfully between the three chosen temperatures across the whole doping range. 
The QMC simulations are run self-consistently shifting the value of $\mu$ to achieve the desired doping. Occupation varies in the full $[-4,4]$ doping range, with steps of $\delta\nu=0.1$. A total number of 100 DMFT iterations are made for each value of $\nu$ and $T$ chosen for the calculation of $S$, and the last $50$ are considered for the determination of the average occupation and chemical potential. This ensures that the steps needed to converge the DMFT run are not counted towards the average. The remaining set of data is then further trimmed by extracting one data point every $\tau$, where this quantity is the integer approximation of the calculated autocorrelation time for the data sample.

The average value of doping and chemical potential are calculated over the resulting set of data points. The error on doping and chemical potential is obtained as mean square deviation over the sample.

An illustration of the difference between the two $\mu(\nu)$ curves is shown in Fig.~\ref{fig:mu_of_nu}. The numerical derivative of the chemical potential with respect to temperature is then calculated as 
\begin{equation}
    \bigg(\dfrac{\partial\mu}{\partial T}\bigg)_{\nu}\approx \dfrac{\mu(T+\Delta T)-\mu(T-\Delta T)}{2\Delta T}
\end{equation}

This is associated to an uncertainty obtained by error propagation as
\begin{equation}
\delta \bigg(\dfrac{\partial\mu}{\partial T}\bigg) = \frac{\sqrt{(\delta \mu(T+\Delta T))^2 + \delta\mu(T-\Delta T) )^2}}{\lvert 2  \Delta T \rvert}
\end{equation}

where we assume the two chemical potentials are uncorrelated, a reasonable assumption after the data trimming process described above.

The entropy is then obtained by numerical integration of the discrete derivative on the doping range, from $+4$ descending to the desired value. This entails $S(\nu=4)=0$, which is a reasonable assumption given the presence of a hard band gap between the fully-occupied flat manifold and the upper dispersive $c$ bands, as shown in Fig.~\ref{fig:akw-panels}. Numerical integration is performed using the Simpson method.

The associated uncertainty is propagated as follows. The numerical calculations involves the integral $S$ of a function $y(x)$, where for simplicity we called $y=\frac{\partial \mu}{\partial T}_{\nu}$ and $x=\nu$. The function is approximated by a discrete set of values $y_{i}$ for each $x_{i}$ with associated uncertainties $\delta y_{i}$ and $\delta x_{i}$. 

In essence, we can express $S$ as a function of the parameter sets $y_{i}$ and $x_{i}$, the distributions of which we can again safely assume to be uncorrelated. The error associated to $S$ is then, through Gaussian propagation at first order,

\begin{equation}
    (\delta S)^{2} = \sum_{i}^{n}\bigg(\dfrac{\partial S}{\partial y_{i}}\delta y_{i}\bigg)^{2} + \bigg(\dfrac{\partial S}{\partial x_{i}}\delta x_{i}\bigg)^{2}
\end{equation}

The contribution to the total uncertainty coming from $y$ essentially comes from the weighted average

\begin{equation}
    y_{n}=\sum_{i}^{n} w_{i}y_{i}
\end{equation}

where $w_{i}$ are the Simpson weights which correspond to $\frac{\partial S}{\partial y_{i}}$ above. The associated MSE is then

\begin{equation}
    \delta y_{n} =\sqrt{\sum_{i}^{n} (w_{i}\delta y_{i})^{2}}
\end{equation}

There is also a contribution to the total uncertainty coming from the $x$ values. Here, we can obtain an estimation by evaluating the integral at $x+\delta x$ and $x-\delta x$ to obtain the numerical derivative $\frac{\partial S}{\partial x_{i}}$, and then expressing the $x$ uncertainty via Gaussian error propagation

\begin{equation}
    \delta x_{n} = \sqrt{\sum_{i}^{n}\dfrac{\partial S}{\partial x_{i}} \delta x_{i}^{2}}
\end{equation}

again assuming the various data points are uncorrelated. Finally, the overall propagated error is calculated as 

\begin{equation}
    \delta S_{n} = \sqrt{\delta x_{n}^{2}+\delta y_{n}^{2}}
\end{equation}

\end{document}